\documentclass[11pt]{article}
\usepackage{jheppub}

\usepackage{color}
\usepackage{amsmath}
\usepackage{pifont}
\usepackage{bbold}

\usepackage{bbm}
\usepackage{verbatim}   
\usepackage{subfigure}
\usepackage{acronym}

\usepackage{amsfonts}
\usepackage{amssymb}
\usepackage{mathrsfs}
\usepackage{graphicx}
\usepackage{multirow}
\usepackage{slashed}
\usepackage[normalem]{ulem}

\graphicspath{{./fig/}}

\newcommand{\kahler}{K\"{a}hler }

  \newcommand{\GeV}{{\mathrm {GeV}}}
   \newcommand{\TeV}{{\mathrm {TeV}}}

\DeclareMathOperator{\z}{\mathbb{Z}_2}

\def\bar#1{\overline{#1}}

\def\inv{^{\raise.15ex\hbox{${\scriptscriptstyle -}$}\kern-.05em 1}}
\def\lbar{{\lower.35ex\hbox{$\mathchar'26$}\mkern-10mu\lambda}} 

\def\to{\rightarrow}

\let\<=\langle
\let\>=\rangle

\let\+=\uparrow

\newcommand{\newc}{\newcommand}
\newc{\gsim}{\lower.7ex\hbox{$\;\stackrel{\textstyle>}{\sim}\;$}}
\newc{\lsim}{\lower.7ex\hbox{$\;\stackrel{\textstyle<}{\sim}\;$}}

\newcommand{\gev}{\ensuremath{\text{ GeV}}}
\newcommand{\tev}{\ensuremath{\text{ TeV}}}

\def\beq{\begin{equation}}
\def\eeq{\end{equation}}
\def\bea{\begin{eqnarray}}
\def\eea{\end{eqnarray}}

\begin{document}

\title{Natural Scherk-Schwarz Theories of the Weak Scale}


\author[a]{Isabel Garc\'ia Garc\'ia,}
\emailAdd{isabel.garciagarcia@physics.ox.ac.uk}

\author[b,c]{Kiel Howe,}
\emailAdd{khowe@fnal.gov}

\author[a,c]{John March-Russell}
\emailAdd{jmr@thphys.ox.ac.uk}

\affiliation[a]{Rudolf Peierls Centre for Theoretical Physics,
University of Oxford,\\
1 Keble Road, Oxford,
OX1 3NP, UK}
\affiliation[b]{Fermilab National Accelerator Laboratory \\ Batavia, IL 60510, USA}
\affiliation[c]{Stanford Institute for Theoretical Physics, Department of Physics,\\
Stanford University, Stanford, CA 94305, USA}

\abstract{
Natural supersymmetric theories of the weak scale are under growing pressure given present LHC constraints, raising the question of whether untuned supersymmetric (SUSY) solutions to the hierarchy problem are possible. In this paper, we explore a class of 5-dimensional natural SUSY theories in which SUSY is broken by the Scherk-Schwarz mechanism. We pedagogically explain how Scherk-Schwarz elegantly solves the traditional problems of 4-dimensional SUSY theories (based on the MSSM and its many variants) that usually result in an unsettling level of fine-tuning. The minimal Scherk-Schwarz set up possesses novel phenomenology, which we briefly outline.
We show that achieving the observed physical Higgs mass motivates extra structure that does not significantly affect the level of tuning (always better than $\sim 10\%$)
and we explore three qualitatively different extensions: the addition of extra matter that couples to the Higgs, an extra $U(1)^\prime$ gauge group under which the Higgs
is charged and an NMSSM-like solution to the Higgs mass problem.
}

\maketitle


\section{Introduction}
\label{sec:intro}

The hierarchy problem, one of the most pressing issues in particle physics,
refers to the statement that the Higgs mass-squared parameter $m_H^2$ is apparently screened from the effects of beyond the Standard Model (SM) physics appearing at scales $\Lambda$ much larger than the scale of electroweak symmetry breaking (EWSB), $v \approx 246 \gev$, despite the lack of a symmetry in the SM enforcing such insensitivity.
Softly-broken Supersymmetry (SUSY) was proposed as a solution to this hierarchy problem and, provided sparticle masses are not too large, is a technically natural solution \cite{Dimopoulos:1981zb}.
However, current experimental measurements constrain the most popular supersymmetric theories, specifically those based on the MSSM and its variants,
to a level of tuning worse than $ \sim 1 \%$ \cite{Gherghetta2013,Arvanitaki2013,Hardy2013,Feng2013,Gherghetta2014,Fan2014},
creating a tension between SUSY and the naturalness principle.

The reasons for the high degree of tuning exhibited by the simplest SUSY theories are manifold.
At tree-level, the Higgs mass-squared gets two types of contributions: a SUSY preserving piece $|\mu|^2$ needed so that
Higgsinos are massive, and a SUSY breaking piece that must be of opposite sign but similar size to the SUSY preserving term.
The two tree-level contributions then need to be tuned against each other in order to achieve a phenomenologically viable electroweak (EW) breaking vacuum expectation value (vev).
The need to generate $\mu$ comparable to the soft masses, linked to the need of writing Higgsino masses, is known as the $\mu$-problem and introduces model building challenges and tuning all ready at tree level.
At 1-loop, radiative contributions to $m_H^2$ proportional to the stop mass-squared and $A$-terms are present and
for these radiative corrections not to be too large, light stops and small $A$-terms are preferred, together with a low SUSY breaking scale.
However, a scenario with light stops, and in particular with stops much lighter than gluinos (experimentally constrained to be heavier),
is difficult to engineer in MSSM-like theories: the stop mass receives log-enhanced radiative corrections proportional to the gluino mass that pulls the former up as the theory is renormalization group evolved from the mediation scale of SUSY breaking down to the IR \cite{Arvanitaki2013}.
A scenario with a light stop and a heavy gluino is therefore typically even more tuned than the case $m_{\tilde t} \sim m_{\tilde g}$.
This tendency of the gluino to pull up the stop and therefore worsen the tuning is dubbed the `gluino sucks' problem.
A natural supersymmetric solution to the hierarchy problem must therefore solve the $\mu$-problem, solve the `gluino sucks' problem (so that parametrically lighter stops can exist without the need for extra tuning) and have naturally small $A$-terms. 
The stringent LHC limits on the first and second generation squarks also pose a problem.  We again must arrange for stops
to be parametrically light compared to the lower generation squarks, a non trivial task in most theories of SUSY breaking
mediation.  Moreover, the observed relatively large physical Higgs mass, $m_h\simeq 125 \gev$, also presents a difficulty: As in the MSSM
such a large mass requires heavy stops ($\gsim {\rm few} \tev$) and large $A$-terms, in direct contradiction to the requirements of naturalness.

In Ref.~\cite{MNSUSY}, a SUSY model of the weak scale was proposed -- Maximally Natural Supersymmetry (MNSUSY) -- that addresses 
the above-mentioned problems.
Excluding the gravitational sector, MNSUSY is a 5-dimensional (5D) supersymmetric theory, with the extra-dimension compactified
on an orbifold with a compactification scale $1/R$ that takes values $\gtrsim 4 \tev$ for a phenomenologically viable model.
The fields propagating in the 5D bulk are the gauge and Higgs sectors together with the 1st and 2nd generation of matter fields,
whereas the 3rd generation remains localized on one of the orbifold branes.
The gauge sector consists purely of the SM gauge group (gauge coupling unification is not realised in the minimal version
of the model, although closely related models in 5D and 6D can realise unification with a precision prediction for $\sin^2\theta_w$ \cite{curly}).  The Higgs sector involves two Higgs supermultiplets although only one gets a non-zero vev, resulting in a SM-like Higgs sector at low energies. SUSY is then broken in the bulk by the Scherk-Schwarz SUSY breaking (SSSB) mechanism with maximal twist \cite{Scherk:1978ta,Scherk:1979zr}, a \emph{non-local} form of SUSY breaking
that links the SUSY breaking scale to the compactification scale $1/R$ while ensuring that SUSY breaking parameters remain (essentially) insensitive to the cutoff $M_*$.
The 5D gauge theory needs a UV completion at a cutoff scale $M_*$ parametrically larger than the compactification scale,
$M_* \lesssim 25 / (\pi R)$ \cite{Muck:2004br,Chivukula:2003kq,curly}.
For example, string implementations of SSSB realize many of the features of the 5D models we consider \cite{Ferrara:1988jx,Kounnas:1988ye,Porrati:1989jk,Antoniadis:1990ew,Antoniadis:1992fh,Ferrara:1994kg,Scrucca:2001ni,Abel:2015oxa}.
We parameterize the dependence of the 5D effective field theory on the details of the UV completion through higher dimensional operators (HDOs) generated at $M_*$.
Effective 5D field theory constructions similar to MNSUSY have been previously considered in the literature
\cite{Antoniadis:1998sd,Delgado:1998qr,Pomarol:1998sd,Delgado:2001si,Delgado:2001xr,Barbieri:2003kn,Barbieri:2002sw,
Barbieri:2002uk,Barbieri:2000vh,Marti:2002ar,Diego:2005mu,Diego:2006py,vonGersdorff:2007kk,Bhattacharyya:2012ct,Larsen:2012rq},
although we believe that this is the first model compatible with both the present stringent LHC constraints on 
sparticle masses and the observed Higgs mass of $m_h\simeq 125 \gev$, while maintaining a low ($\sim 30\%$) level of tuning.

As a result of the SSSB mechanism, bulk fields are affected by the breaking of SUSY at tree-level, whereas brane-localized fields only pick up masses
at 1-loop from radiative corrections involving bulk fields.
The tree-level spectrum is such that tree-level Higgsino masses are $\approx 1/(2R)$, whereas the Higgs scalar mass-squared vanishes, solving the $\mu$-problem.
Similarly, gaugino masses are of Dirac nature and of size $\approx 1/(2R)$, whereas the stop, being brane-localized, remains massless at tree-level.
At 1-loop, stop masses are generated with the main contribution coming from the gluino sector. However, due to the non-local nature of the SSSB mechanism
all SUSY breaking quantities are only sensitive to scales up to $\sim 1/R$, and the stops remain naturally parametrically lighter than the gluino
(typically $m_{\tilde t} \sim 1/(10R) \sim m_{\tilde g} / 5$), solving the `gluino sucks' problem.
1st and 2nd generation sfermions also pick up tree level masses equal to $1/(2R)$, automatically making 1st and 2nd generation squarks parametrically heavier than stops, so implementing the `natural SUSY' spectrum \cite{Dimopoulos:1995mi,Pomarol:1995xc,Cohen:1996vb}. 

An extra feature of the theory that arises in the particular case of SSSB with maximal twist is the presence of an accidental $U(1)_R$ $R$-symmetry.
Among other features, this accidental $U(1)_R$ forbids the presence of $A$-terms -- another ingredient that helps minimise tuning -- and greatly
ameliorates FCNC and rare process constraints on the sparticle spectrum \cite{MNSUSYflavor}.
It is worth emphasising that the SSSB mechanism with maximal twist differs qualitatively from the case of general (non-zero but non-maximal) twist in that the former
is a symmetry enhanced point, since only in the limit of maximal twist does the accidental $U(1)_R$ arise:
SSSB with maximal twist \emph{is} a special point in terms of symmetries\footnote{We comment on the effect gravitational interactions may have on this $U(1)_R$
symmetry in Appendix \ref{app:MaximalRadionMed}.}. 

An important additional ingredient of the theory is an extra SUSY breaking sector from the requirement of tuning the 4D cosmological constant (CC) to nearly zero, which after radius stabilization takes place tends to be
of order $\sim 1/(\pi R)^4$ and negative
\cite{Rohm:1983aq,Ponton:2001hq,vonGersdorff:2003rq,Rattazzi:2003rj,vonGersdorff:2005ce,Dudas:2005vna,Gross:2008he}.
The important effects on the low-energy spectrum and EWSB are parameterized by couplings to a SM-singlet brane-localised chiral superfield $X$, whose $F$-term gets a vev $F_X \sim 1/(\pi R)^2$. In particular regarding EWSB, a positive 1-loop contribution to $m_H^2$ arises from the bulk EW sector, and a negative piece from
the stop sector of parametrically similar size is present at 2-loop order.
However, these two contributions are not enough to trigger EWSB and an extra contribution is required. This extra piece is naturally present from
HDOs involving both $X$ and the Higgs and/or stop supermultiplets. Coefficients of $\mathcal{O}(1)$ for these HDOs trigger successful EWSB resulting in a very mild ($\sim 30\%$) level of fine-tuning, much better than 4D SUSY theories built around the MSSM.

Unfortunately, all the ingredients that allow for a natural theory of EWSB result in a physical mass for the Higgs that is too low compared to the experimental measurement $m_h \simeq 125 \gev$.
This requires the addition of some extra structure that would contribute to the Higgs quartic coupling and raise $m_h$ to its observed value.
In this respect, several options are possible that do not significantly alter the physics of EWSB.
We explore three different possibilities: (i) a generation of brane-localized vector-like leptons that couple to the Higgs with $\mathcal{O}(1)$ Yukawa couplings, (ii) extra gauge structure in the bulk,
and (iii) a case where both Higgs doublets get non-zero vev's and a brane-localized SM-singlet chiral superfield is added that allows for an extra contribution to the physical Higgs mass (similar to the situation in the NMSSM).

While we focus on one specific realization of a natural 5D SUSY theory in this work, many of the mechanisms and results regarding EWSB and Higgs properties can be applied to similar models with different choices of field localization \cite{MNSUSYflavor,Delgado:2001si,Delgado:2001xr,Diego:2005mu,Diego:2006py}, quasi-localized matter \cite{vonGersdorff:2007kk,Bhattacharyya:2012ct,Barbieri:2003kn,Barbieri:2002sw,Barbieri:2002uk,Barbieri:2000vh,Marti:2002ar}, bulk curvature \cite{Katz:2006mva,Gherghetta:2000kr}, and even extended symmetry structures such as those found in Folded SUSY models~\cite{Burdman:2006tz,Cohen:2015gaa}.

The structure of the paper is as follows.
In Section \ref{sec:toy}, we consider a toy model that illustrates the basics of the SSSB mechanism and the role of the different ingredients present
in the minimal model.
Section \ref{sec:model} contains a realistic description of the minimal theory, featuring all necessary ingredients, as well as a discussion of EWSB and the status
of the physical Higgs mass.
In Section \ref{sec:extensions}, we consider several extensions of the model that allow for a physical Higgs mass consistent with observations and low fine-tuning.
The phenomenology of the different versions of the model considered is discussed in Section~\ref{sec:pheno}.
Finally, Section~\ref{sec:conclusions} contains our conclusions and an appendix helps clarify the physics of SSSB with maximal twist
from the point of view of radion mediation and supergravity (SUGRA).

\section{The basics of Scherk-Schwarz SUSY breaking}
\label{sec:toy}

\begin{figure}[h!]
\begin{centering}
\includegraphics[scale=0.35]{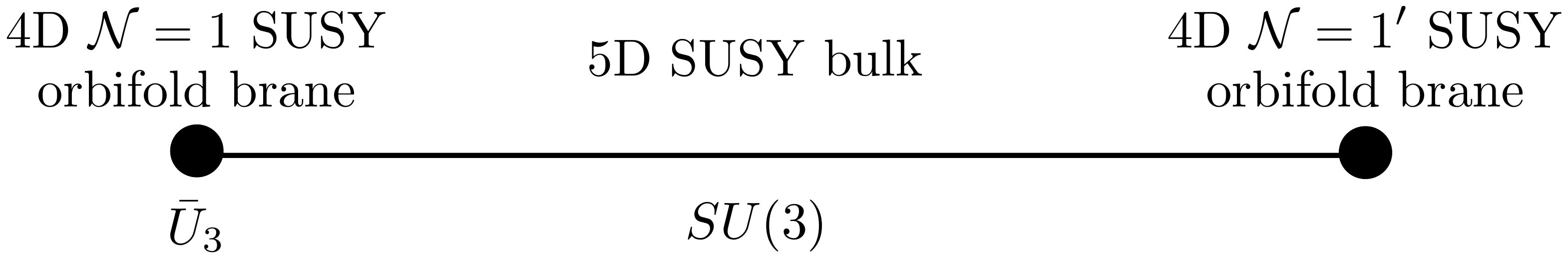}
\caption{\label{fig:MoreMinimalGeography} 
	Schematic geography of the toy model containing bulk $SU(3)$ interactions and a brane-localized right-handed top superfield $\bar{U}_3$.}
\end{centering}
\end{figure}

For simplicity of presentation, and after a very brief discussion in Section~\ref{sec:5DSUSY} of 5D SUSY and the generalities of SSSB,
this Section focuses on the physics of a toy model containing just the right-handed top superfield $\bar{U}_3$ (on shell: $\bar{U}_3 = ({\tilde {\bar u}}_3, \bar u_3)$,
with $\bar u_3$ a 2-component Weyl fermion)
localized on the orbifold $y=0$ brane and the $SU(3)$ color gauge group in the bulk, as depicted in Figure~\ref{fig:MoreMinimalGeography}.
Although this simplified model is inconsistent in several ways as a stand-alone theory, it forms part of the final and fully consistent picture that will be described in Section~\ref{sec:model}
and contains the minimal ingredients that will help us illustrate some of the most important features of maximal SSSB.  Specifically, these include
(i) large hierarchies in soft SUSY breaking masses generated at tree level by the 5D geography,
(ii) the very soft nature of loop communication of SSSB,
and (iii) the potential importance of extra SUSY breaking sectors associated with radius stabilisation.

\subsection{5D SUSY with Scherk-Schwarz breaking} 
\label{sec:5DSUSY}

We start with a short review of the relevant aspects of 5D SUSY and SSSB.  
Readers already familiar with this material are encouraged to jump to Section~\ref{sec:bulkstates}.

A realistic description of the world arising from an extra-dimensional theory can be constructed provided the extra dimensions are compactified.
From a 4D perspective, fields that propagate along the bulk of the extra dimension are equivalent, upon compactification, to an infinite tower
of fields of ever increasing mass---the Kaluza-Klein (KK) excitations---with a mass gap between different modes set by the compactification scale.
For example, if an extra spatial dimension is compactified on a circle $S^1$ of radius $R$,
KK modes have masses that are multiples of the compactification scale $1/R$.
Nevertheless, to obtain chiral fermions in the low energy theory the extra dimension needs to be compactified not
on a circle but on an `orbifold' where discrete identification(s) of the extra-dimensional coordinate, such as $y\sim -y$ with associated
fixed points (here $y=0,\pi R$), imply the presence of 4D branes violating the higher-dimensional bulk Lorentz symmetry.
For example, an extra spatial dimension compactified on an orbifold $S^1 / \mathbb{Z}_2$ has physical length $\pi R$,
with the two orbifold fixed-points at $y=0, \pi R$ corresponding to 4D branes where fields and interactions may be localized.
This $\mathbb{Z}_2$ orbifold action is then extended to an action on the bulk fields; e.g. for a bulk scalar the identification
$\phi(x^\mu,y) = \pm \phi(x^\mu,-y)$ can be made, which translates to Neumann (for $+$) or Dirichlet (for $-$)
boundary conditions (bc's) for the bulk field at the fixed points. 

The minimal representation of SUSY in 5D corresponds to
$\mathcal{N}=2$ extended SUSY from a 4D perspective, i.e. the theory has 8 rather than 4 supercharges. The $\mathcal{N} = 2$ SUSY can be
expressed in $\mathcal{N} = 1$ superfield notation \cite{ArkaniHamed:2001tb,Marti:2001iw,Hebecker:2001ke}.  An $\mathcal{N} = 2$ vector superfield $\mathbb{V}^a$ may be
written in terms of one $\mathcal{N} = 1$ vector superfield $V^a$ plus one $\mathcal{N} = 1$ chiral superfield $\chi^a$, both transforming under the adjoint representation of the corresponding gauge group (on shell: $\mathbb{V}^a = \{ V^a, \chi^a\}$ with $V^a = ( V^a_\mu, \lambda^a )$ and $\chi^a = ( \bar \sigma^a, \bar \lambda^a )$ respectively, with the complex adjoint scalar $\bar \sigma^a$ containing $V^a_5$).
A hypermultiplet $\mathbb{H}$, the $\mathcal{N} = 2$ generalization of the 
familiar $\mathcal{N} = 1$ chiral multiplet, can be described in terms of two $\mathcal{N} = 1$ chiral superfields
with on shell content $\mathbb{H} = \{ \Phi, \Phi^c \}$ with $\Phi = ( \phi, \psi )$ and $\Phi^c = ( \phi^c, \psi^c )$. 
Notice $\lambda^a$ and $\bar \lambda^a$, as well as $\psi$ and $\psi^c$, denote two independent 2-component Weyl fermions. 

After the extra dimension is compactified on an $S^1 / \mathbb{Z}_2$ orbifold, a minimal $\mathcal{N} = 1$ 4D
SUSY (4 supercharges) survives in the 4D effective theory, as the other supercharges and the corresponding extra bulk  superpartners are removed by the action of the orbifold bc's. While the full 5D SUSY survives locally in the bulk, 4D $\mathcal{N}=1$ supersymmetric field content and interactions may be localized on the two 4D orbifold branes. The bulk  $\mathcal{N} = 2$ SUSY theory possesses an enhanced $R$-symmetry, $SU(2)_R$, 
which is also broken to the usual $U(1)_R$ of minimal 4D $\mathcal{N} = 1$ SUSY
upon compactification on $S^1 / \mathbb{Z}_2$. If the compactification involves further orbifold actions or twisted boundary condtions, all of the 4D SUSY can be broken. For example, an $S^1 / (\mathbb{Z}_2 \times \mathbb{Z}^\prime_2)$ orbifold has two inequivalent fixed points (and thus branes) and can, depending on the choice of field parities under the $\mathbb{Z}_2 \times \mathbb{Z}^\prime_2$ actions, break SUSY completely in the IR theory \cite{Barbieri:2001dm}. 

As explained in detail in e.g. \cite{Pomarol:1998sd,Quiros:2003gg}, SSSB consists in breaking SUSY \emph{non-locally} by imposing on the 5D fields a non-trivial periodicity condition under translation around the 5th dimension (a more detailed discussion can also be found in Appendix~\ref{app:bcpicture}). The twist is a rotation in the $SU(2)_R$ group of the 5D SUSY, which acts differently on the bosonic and fermionic components of a given supermultiplet.  The twist does not commute with the SUSY preserved by the original orbifold and breaks the remaining $\mathcal{N} = 1$ 4D SUSY.  Crucially, the twist is only defined \emph{globally}, with any local physical observable
being unaffected by the twist, and SUSY is unbroken \emph{locally}.  Intuitively, only Feynman diagrams that stretch all the way across the (finite sized) extra dimension are sensitive to SUSY breaking, and such diagrams are finite and free from (even logarithmic) sensitivity to UV scales above $1/R$ \cite{Antoniadis:1997zg,Barbieri:2001dm,Delgado:2001ex,Contino:2001gz,Weiner:2001ui,Kim:2001re,Puchwein:2003jq}.
This non-local nature of SSSB is the fundamental reason why SUSY breaking parameters are UV insensitive, for above the scale $1/R$ the theory \emph{is} supersymmetric.

In this work we will exclusively focus on the case where the Scherk-Schwarz twist is maximal, in which case the spectrum of the theory is equivalent to compactification on an $S^1 / (\mathbb{Z}_2 \times \mathbb{Z}^\prime_2)$ orbifold with a particular choice of $\mathbb{Z}_2 \times \mathbb{Z}^\prime_2$ field parities\footnote{We choose the fundamental interval to be $\pi R$, corresponding in the case of the $\mathbb{Z}_2 \times \mathbb{Z}^\prime_2$ orbifold of a covering $S^1$ of radius $2R$. Parities under the two $\mathbb{Z}_2$ symmetries given by $(+,+)$, $(+,-)$, $(-,+)$ and $(-,-)$ correspond to a KK spectrum of modes given by $m_n = n/R$, $(2n+1)/2R$, $(2n+1)/2R$ and $(n+1)/R$ respectively}. 
The 5D theory always locally preserves at least $\mathcal{N} = 1$ SUSY (this is true at the fixed points; in the bulk,
$\mathcal{N} = 2$ SUSY is locally preserved), but is broken down to \emph{inequivalent and incompatible}
$\mathcal{N}=1$ and $\mathcal{N}=1'$ SUSYs on the two orbifold fixed-points separated by a physical distance $\pi R$.  Thus
the maximal SSSB mechanism completely breaks SUSY in the effective 4D theory below the scale $\sim 1/(\pi R)$, with the
transition from the 5D theory to the 4D theory happening at the same scale as that at which SUSY is fully broken.  We emphasise that
there is \emph{no} regime where the theory is a softly broken 4D SUSY theory. In this respect, as in many others, 4D theories based
upon SSSB radically differ from the MSSM and its variants.
Finally, as mentioned in Section~\ref{sec:intro}, the case of SSSB with maximal twist is a symmetry enhanced point,
for the IR theory possesses a $U(1)_R$ $R$-symmetry, the details of which will be discussed in Section~\ref{sec:U(1)R}.

In the case of non-maximal twist, the spectrum for a single bulk hypermultiplet is always such that a fermion component stays massless
whereas the scalars pick up SUSY breaking masses at tree-level from the twist. However, for the Higgs multiplets, a spectrum with massless scalars and massive Higgsinos is desired. When \emph{both} $H_u$ and $H_d$ hypermultiplets are present in the bulk, a SUSY-preserving even-parity bulk mass can be introduced to modify the spectrum, which can be equivalently described as a SUSY-preserving twist by an $SU(2)_H$ symmetry under which the two Higgs hypermultiplets transform as doublets \cite{Pomarol:1998sd,Quiros:2003gg}. This twist can be chosen to leave a scalar 0-mode massless while the Higgsinos are lifted. Therefore at non-maximal $SU(2)_R$ twist, the presence of a tree-level 0-mode in the Higgs sector may be regarded as a result of tuning between the $SU(2)_R$ and $SU(2)_H$ twists. However, at maximal $SU(2)_R$ twist the situation is special again: the maximal $SU(2)_H$ twist corresponds simultaneously to massless scalars and an enhanced $U(1)_{H_d}\times U(1)_{H_u}$ global symmetry (the $U(1)_{H_u}$ symmetry will be broken by Yukawa interactions). This is manifest in the $S^1 / (\mathbb{Z}_2 \times \mathbb{Z}^\prime_2)$ orbifold language, where boundary conditions giving a massless scalar are a discrete choice, as discussed in Section~\ref{sec:fullmodel}.

\subsection{Bulk States}
\label{sec:bulkstates}

Going back to our toy model, depicted in Figure~\ref{fig:MoreMinimalGeography}, we now proceed to illustrate the physics of the SSSB mechanism.
The field content of this toy model is an $\mathcal{N}=2$ vector supermultiplet that propagates in the bulk and transforms under the adjoint representation of $SU(3)$
together with a 4D $\mathcal{N}=1$ chiral superfield $\bar U_3$ localized in the $y=0$ brane.
The parities under the two $\mathbb{Z}_2$ symmetries of the bulk field are chosen such that the gauge field has a massless KK 0-mode
(to be identified with the 4D gluon), and can be summarized as follows: $(+,+)$ for $V^a_\mu$, $(+,-)$ for $\lambda^a$, $(-,+)$ for $\bar \lambda^a$ and $(-,-)$ for $\bar \sigma^a$.  Since these parities lead to the lightest KK mode gluinos being massive, but keep the gluons massless, SUSY is, as advertised,
fully broken.  From a phenomenological point of view, we will be interested in the lightest KK modes of the 5D states.
The gauginos $\lambda^a$ and $\bar \lambda^a$ obtain purely Dirac masses, and their lightest modes form a Dirac gluino with mass $M_3 = 1/(2R)$.
The Dirac nature of the gaugino masses is a consequence of the $U(1)_R$ symmetry preserved by the maximal Scherk-Schwarz twist.
The adjoint scalars and KK excitations of the 4D vector fields begin to appear at $1/R$.

The bulk 5D gauge interactions are non-renormalizable and a 5D gauge theory is necessarily an effective theory valid up to a scale $M_*$. The 5D gauge couplings $g_{I,5}$ are dimensionful, and the 4D gauge couplings at the matching
scale $\mu\simeq 1/R$ are given by $1/g_{I,4}^2 = \pi R/g_{I,5}^2$ up to small brane-kinetic-term corrections, and the 5D perturbative unitarity bound on $g_3$ requires $M_* \pi R \lesssim 25$ \cite{Muck:2004br,Chivukula:2003kq,curly}, corresponding $M_* \sim 30-100~\TeV$ for the TeV-scale radii we consider.  We will parameterise other potential strong-coupling UV effects with $\tilde g = g_{3,5} \sqrt{M_*} = g_{3,4} \sqrt{M_* \pi R} \approx \sqrt{M_* \pi R}$ as the dimensionless strong coupling parameter at the scale $M_*$, so that for example the 5D scalar Lagrangian can be expressed as a function of the brane fields $\phi$ and bulk fields $\Phi$ with $\mathcal{O}(1)$ coefficients\footnote{The relative factor of $\sim \pi$ in the normalization of brane and bulk strong coupling expansion results from the different phase space available to bulk and brane states. In the thin-brane limit, NDA gives the relative 4D ($16\pi^2$) and 5D  ($24\pi^3$) loop factors \cite{Chacko:1999hg,Luty:1997fk,Georgi:1992dw}. Equivalently, in the fat-brane picture the factor of $\sim \pi$ is reproduced in the limit that the brane width appproaches the fundamental lenght $l_b\sim \pi/M_*$.} \cite{Chacko:1999hg},
\beq
\mathcal{L} = \frac{M_*^5}{\tilde{g}^2}\mathcal{L}_5\left(\frac{\partial}{M_*}, \frac{\tilde g \Phi}{M_*^{3/2}}\right) + \frac{3 \pi}{2}\frac{M_*^4}{\tilde{g}^2}\delta(y)\mathcal{L}_0\left(\frac{\partial}{M_*}, \frac{\tilde g \Phi}{M_*^{3/2}}, \frac{\tilde g \phi}{\sqrt{3\pi/2}M_*}\right).
\label{eq:NDA}
\eeq

For example, SUSY-preserving brane-localized kinetic terms are the lowest dimension derivative operators that can be generated.
For bulk vector fields they appear in the superpotential in the form
\beq
\Delta W \sim \frac{3\pi / 2}{M_*} W_\alpha  W^\alpha \delta(y),
\label{eq:BraneLocalizedKinetic}
\eeq
where $W_\alpha$ refers to the superfield strength of the $\mathcal{N}=1$ vector supermultiplet. The choice of bulk bc's guarantees the existence of a 0-mode for the gauge bosons and the absence of 0-modes for the gauginos and extra $\mathcal{N}=2$ scalar superpartners, but the spectrum of the bulk superpartners and KK excitations can be perturbed by these operators, giving $\sim10-20\%$ deviations for $M_* \pi R \sim 25$.  Brane-localized operators with transverse derivatives have comparable effects \cite{Lewandowski:2001qp}.
This perturbation of the spectrum can be important for the phenomenology of the heavy bulk superpartners and KK modes. However, we will mostly be focused on the spectrum and phenomenology of brane-localized superpartners and bulk 0-modes,
and for these purposes the effects of these perturbations on the bulk KK spectrum can be safely disregarded (for smaller hierarchies $M_* \pi R \ll 25$ these perturbations can be large enough to affect EWSB, but we do not consider this limit since it corresponds to the breakdown of predictivity in the 5D theory). As we will see in Section~\ref{sec:Fx}, another set of
HDOs will parameterize a more important effect on EWSB and the spectrum of brane-localized states.

\subsection{Brane-Localized States}
\label{sec:oneloopSSSB}

Because the right-handed top is localized on the $y=0$ brane, it has no KK excitations and need only form part of a supermultiplet, $\bar U_3$,
consistent with the $\mathcal{N}=1$ SUSY preserved on the brane (just as in the MSSM).
At tree level, locality in 5D protects the $\bar U_3$ multiplet from the breaking of this $\mathcal{N}=1$ SUSY by the  incompatible
$\mathcal{N}=1'$ SUSY on the $y=\pi R$ brane, and a SUSY-breaking mass for the scalar will only be generated by
SSSB at loop level.

The 1-loop contributions to the scalar mass involve SUSY breaking bulk loops of the gluons and gluinos propagating between the $y=0$ and $y=\pi R$ brane. The 1-loop contribution gives a \emph{finite} positive mass squared \cite{Antoniadis:1998sd}, 
\begin{gather}
	{\tilde m}_{U_3}^2 =  \frac{7 \zeta(3)}{16 \pi^4}\frac{4 g_3^2}{3}\frac{1}{R^2} \approx \left(\frac{1}{10 R}\right)^2 \approx \left(\frac{1}{5} M_3\right)^2.
\label{eq:softmassg3}
\end{gather}
The gluino is naturally five times heavier than the stop! In MSSM-like models, even a small amount of running from the messenger scale pulls the stop mass to within a factor of two of the gluino mass unless the parameters are specially tuned to give a hierarchy \cite{Arvanitaki2013}, and the comparatively large built-in gluino-stop hierarchy in SSSB is attractive.  The softness of SSSB arises because loops communicating SUSY breaking must propagate between both branes and are exponentially suppressed at large 4-momenta $|p_4| > 1/(\pi R)$ \cite{Antoniadis:1998sd}, giving an effective messenger scale for SUSY breaking of $\sim 1/(\pi R)$.  

If only SSSB effects are present, then the spectrum for both the brane-localised superpartners and the bulk superpartners is very predictive, with the dominant effects determined completely by the scale $1/R$ and the choices of bc's. However, there is a generic possibility for additional sources of SUSY breaking from the dynamics of radius stabilisation which we discuss in Section~\ref{sec:Fx}.
While this will have a small effect on the spectrum of bulk superpartners like the $SU(3)$ gauginos, we will find it can have an $\mathcal{O}(1)$ effect on the spectrum of brane-localized sfermions.

\subsection{SUSY Breaking from Radius Stabilisation}
\label{sec:Fx}

A phenomenologically consistent theory requires that the scale of the compactified 5th dimension $\sim 1/R\sim\TeV \ll M_*$ must be dynamically stabilised, otherwise a massless radion mode with excluded couplings would exist in the spectrum.  In pure SSSB, the only ingredients in the radion potential are supersymmetric brane tensions, a supersymmetric 5D CC, and SUSY-breaking Casimir energies induced by the Scherk-Schwarz bc's.  Although these ingredients can stabilise the radion, they generically do so at a non-vanishing (and negative) value of the 4D CC, and additional sources of SUSY-breaking contributions to the potential are needed to lift the minimum to a vanishing 4D CC \cite{Ponton:2001hq,vonGersdorff:2003rq,Rattazzi:2003rj,vonGersdorff:2005ce,Dudas:2005vna,Gross:2008he}.

For example, if the brane tensions, bulk CC, and the Casimir energy of the minimal bulk matter content are the only ingredients in the stabilisation sector, then the brane and bulk tensions must break SUSY to stabilise the radius with vanishing 4D CC \cite{vonGersdorff:2003rq}. With non-minimal field content the radius can have meta-stable minima with SUSY preserving brane tensions, for example if the theory contains additional quasi-localized states \cite{vonGersdorff:2003rq},
but generic stable minima with vanishing 4D CC require SUSY breaking tensions \cite{Rattazzi:2003rj}.
If the radius is stabilised by additional tree-level dynamics for bulk fields \cite{Goldberger:1999uk}, then the brane dynamics also generically lead to brane-localized $F$-terms. 

While it is not necessary to fully specify the radion stabilisation dynamics to study the properties of the SM fields and their superpartners in SSSB models, we find it is important to parameterise the effects of the additional sources of SUSY breaking which may be present to cancel the 4D CC. We thus study the effects of a brane-localized SUSY-breaking tension, which we parameterise by a hidden sector field $X$ with $F$-term $F_X$.  The Casimir energy of the bulk gravitational, gauge, and matter states $V_C \sim 1/(\pi R)^4$ \cite{vonGersdorff:2003rq} sets the typical scale for contributions to the radion potential,
and therefore sets a typical scale of $F_X  \sim 1/(\pi R)^2$ to cancel against other contributions and give a vanishing 4D CC. 

Operators coupling the SM superpartners to $X$ can generate additional soft SUSY breaking masses beyond those originating
at tree and loop-level from the Scherk-Schwarz bc's. 
For example, if $X$ is localized on the same brane as $\bar U_3$, then in the strong coupling expansion of Eq.~(\ref{eq:NDA}) there are dimension six \kahler operators coupling these states,
\begin{equation}
\mathcal{K}  \supset   \delta(y) c_3 \left(\frac{2\tilde{g}^2}{3\pi M_*^2} \right)\left( X^\dagger X  \bar U_3^\dagger \bar U_3 \right) ,
\label{eq:stopHDO}
\end{equation}
with $c_3$ an $\mathcal{O}(1)$ coefficient. This gives a SUSY breaking scalar mass of size
\begin{equation}
	\Delta {\tilde m}_{U_3}^2 \approx c_3 f_X^2  \left(\frac{25}{M_* \pi R}\right) \times \left(\frac{1}{30 R}\right)^2.
\label{eq:stopmassHDO}
\end{equation}
where the dimensionless $\mathcal{O}(1)$ quantity $f_X$ is defined as $ F_X  \equiv f_X / (\pi R)^2$.
This contribution to the SUSY breaking mass of ${\tilde {\bar u}}_3$ can be comparable to the 1-loop minimal Scherk-Schwarz contribution Eq.~(\ref{eq:softmassg3}).

Therefore the on-brane spectrum can be perturbed by contributions from $F_X$ comparable to the 1-loop SSSB masses, leading to a prediction for the overall scale of masses of on-brane states with $\mathcal{O}(1)$ uncertainty. This has important phenomenological consequences for the production and decay of brane-localized superpartners, and we will also find that the extra contribution to the stop mass can be important to radiatively drive EWSB.
(SUSY breaking effects from the radius stabilisation sector can also be communicated by anomaly mediation,  but as discussed in Appendix~\ref{app:MaximalRadionMed} these are negligible compared to the direct $F_X$ terms and the loop-level SSSB effects.)

\section{Minimal Model}
\label{sec:model}

The discussion so far of the $SU(3)$ gauge multiplet and right-handed top chiral supermultiplet $\bar U_3$ has illustrated some key features of SSSB.
The gauge multiplet propagates in the bulk, and its $\mathcal{N}=2$ superpartners obtain large tree-level SUSY breaking masses from the maximal SUSY breaking bc's, with gluinos obtaining a Dirac mass $M_3=1/(2R)$.
The $\bar U_3$ multiplet is localized on the brane, and the stop squark obtains a mass from SUSY breaking gluino bulk loops $\tilde m^2_{U_3} \approx 1/(10 R)^2$.
The stop can also obtain a comparable contribution to its mass from additional sources of SUSY breaking $F_X$ associated with cancelling the 4D CC.
In this section, we extend the previous toy model to a realistic theory that includes the absolute minimal ingredients. This model fails to give a $125~\GeV$ Higgs mass in the parameter regime of low tuning, and we discuss extensions motivated by this problem in Section~\ref{sec:extensions}.

\subsection{Full matter content and gauge interactions}
\label{sec:fullmodel}

\begin{figure}[h!]
\begin{centering}
\includegraphics[scale=0.35]{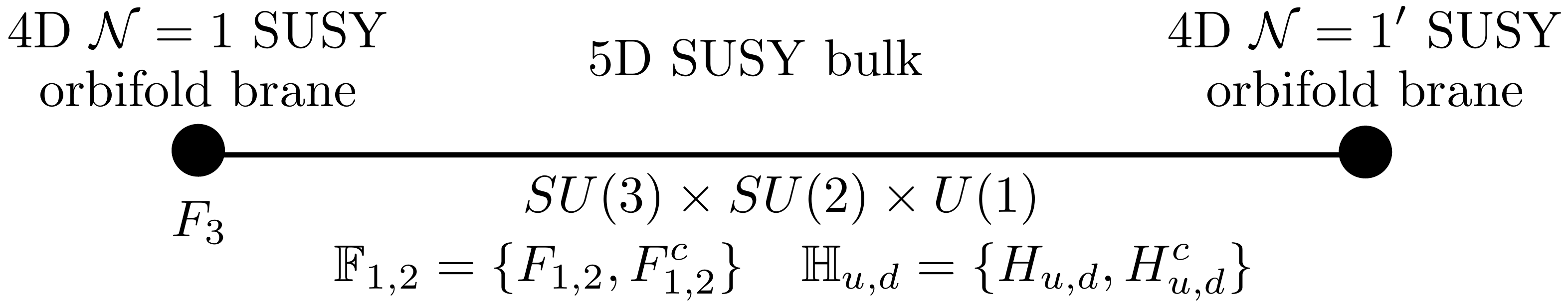}
\caption{\label{fig:MinimalGeography} Schematic geography of natural spectrum embedded in a 5D Scherk-Schwarz model.
Gauge and Higgs sectors, together with the 1st and 2nd generation of matter, propagate in the extra dimensional bulk.
Each SM state is accompanied by a full 5D SUSY multiplet and KK excitations, and SUSY breaking is felt by these states at tree level from the Scherk-Schwarz bc's. The 3rd generation states are localized on the brane at $y=0$ and fill out multiplets of the locally preserved $\mathcal{N}=1$ SUSY; SUSY breaking is communicated to these states at loop level by their interactions with the 5D gauge and Higgs fields.
$F_i$ referes to the usual $\mathcal{N}=1$ chiral supermultiplets needed for each generation, i.e.
$F_i = Q_i, \bar U_i, \bar D_i, L_i, \bar E_i$ ($i=1,2,3$).}
\end{centering}
\end{figure}

The results discussed in Section~\ref{sec:toy} can be easily generalized in order to build a model that contains the full SM matter and gauge interactions at low energies --
the minimal ingredients of such model are depicted in Figure~\ref{fig:MinimalGeography}.
The gauge and Higgs sectors propagate in the bulk, with the former extended to include the full SM gauge group $SU(3) \times SU(2) \times U(1)$. As before, gauginos get Dirac masses at tree-level of size $M_{1,2,3} = 1/(2R)$ from maximal SSSB.  Regarding matter content,  the 1st and 2nd generations also propagate in the 5D bulk, and therefore five hypermultiplets for each of the two families are present:
$\mathbb{F}_i = \{ F_i, F_i^c \}$ ($i=1,2$), where $F_i$ refers to the usual $\mathcal{N}=1$ chiral superfields present in the MSSM
($F_i=Q_i, \bar U_i, \bar D_i, L_i, \bar E_i$).
SSSB bc's are such that chiral fermions (the 1st and 2nd generation SM quarks and leptons) remain in the low energy theory whereas sfermions of the 1st and 2nd generation get tree-level masses of size $1/(2R)$ (degenerate with gauginos, and, as we will soon argue, Higgsinos)
and the first conjugate fermion partner appears paired with the first KK excitation of the SM fermion with a Dirac mass of $1/R$.
Because the first two generations appear as bulk states, the spectrum of superpartners and KK excitations can be perturbed at the $\lesssim 10\%$ level by brane-localized kinetic terms without significant consequences, just as described for the gauge multiplets.


The Higgs sector includes two hypermultiplets $\mathbb{H}_{u,d} = \{ H_{u,d}, H_{u,d}^c\}$. In the simplest version of our model, only $H_u$ will obtain a non-zero vev, which automatically results in a SM-like Higgs sector at low energies\footnote{We discuss the structure of the Yukawa interactions in Section~\ref{sec:yukawas}.}
and $H_d$ is therefore left as an inert Higgs doublet\footnote{In Section~\ref{sec:singlet} we argue that one of variant models that accommodates the
Higgs mass, $m_h \simeq 125~\GeV$, involves an NMSSM-like structure with both $H_u$ and $H_d$ acquiring
vevs.}. We have chosen a model with both $\mathbb{H}_u$ and $\mathbb{H}_d$ Higgs bulk hypermultiplets with opposite hypercharge only to avoid a brane-localized Fayet-Iliopoulos (FI) term generated at 1-loop, which induces a UV-sensitive 5D odd-parity mass term for those hypermultiplets charged under $U(1)_Y$ \cite{Ghilencea:2001bw,Barbieri:2002ic,Marti:2002ar}. 
At tree-level, an accidental global $U(1)_{H_d}$ symmetry (or $\mathbb{Z}_k$) acting on $\mathbb{H}_d$ is present, which can be chosen
to remain exact and forbids a $\mu$-term.
Despite the
absence of $\mu$-terms, Higgsinos, similar to gauginos, get a Dirac tree-level mass of size $1/(2R)$ due to the SSSB bc's, whereas \emph{the contribution to the mass-squared parameters of the scalar components of the Higgs multiplets is strictly zero at tree-level}.  This elegantly solves the $\mu$-problem! A few 4D mechanisms for heavy Higgsinos without large contributions to the scalar masses exist \cite{Nelson:2015cea,Cohen:2015ala,Perez:2008ngy}, but SSSB with maximal twist provides a qualitatively different 5D realisation protected by the $U(1)_{H_d}$ and $U(1)_R$ symmetries.  Of course the radiative tuning of the Higgs soft mass is irreducible, and we find in Section~\ref{sec:EWSB}  that at loop level the Higgs 0-modes do obtain \emph{finite} SUSY breaking masses.

The bc's for all the fields that propagate in the extra dimensional bulk are summarised in Table~\ref{tab:BoundaryConditions}
and are chosen to be consistent with the two inequivalent and incompatible $\mathcal{N}=1$ and $\mathcal{N}=1'$ SUSYs that are preserved at, respectively, $y=0$ and $y=\pi R$.

\begin{table}[h!]\centering
\begin{tabular}{|c|c|c|c|c|}
\cline{1-5}
\vbox to2.5ex{\vspace{1pt}\vfil\hbox to20ex{\hfil $$\hfil}} & 
\vbox to2.5ex{\vspace{1pt}\vfil\hbox to11ex{\hfil $(+,+)$ \hfil}} & 
\vbox to2.5ex{\vspace{1pt}\vfil\hbox to11ex{\hfil $(+,-)$\hfil}} & 
\vbox to2.5ex{\vspace{1pt}\vfil\hbox to11ex{\hfil $(-,+)$\hfil}} & 
\vbox to2.5ex{\vspace{1pt}\vfil\hbox to11ex{\hfil $(-,-)$\hfil}} \\

\cline{1-5}
\vbox to2.5ex{\vspace{1pt}\vfil\hbox to20ex{\hfil $\mathbb{V}^a = \{ V^a, \chi^a \}$\hfil}} & 
\vbox to2.5ex{\vspace{1pt}\vfil\hbox to11ex{\hfil $V^a_\mu$\hfil}} & 
\vbox to2.5ex{\vspace{1pt}\vfil\hbox to11ex{\hfil $\lambda^a $\hfil}} &  
\vbox to2.5ex{\vspace{1pt}\vfil\hbox to11ex{\hfil $\bar{\lambda}^a $\hfil}} & 
\vbox to2.5ex{\vspace{1pt}\vfil\hbox to11ex{\hfil $\bar{\sigma}^a $ \hfil}} \\

\cline{1-5}
\vbox to2.5ex{\vspace{1pt}\vfil\hbox to20ex{\hfil $\mathbb{H}_{u,d} = \{ H_{u,d}, H_{u,d}^c \}$\hfil}} & 
\vbox to2.5ex{\vspace{1pt}\vfil\hbox to11ex{\hfil $ h_{u,d} $\hfil}} & 
\vbox to2.5ex{\vspace{1pt}\vfil\hbox to11ex{\hfil $ \tilde h_{u,d} $\hfil}} &  
\vbox to2.5ex{\vspace{1pt}\vfil\hbox to11ex{\hfil $ \tilde h_{u,d}^c $\hfil}} & 
\vbox to2.5ex{\vspace{1pt}\vfil\hbox to11ex{\hfil $ h^c_{u,d} $ \hfil}} \\

\cline{1-5}
\vbox to2.5ex{\vspace{1pt}\vfil\hbox to20ex{\hfil $\mathbb{F}_{1,2} = \{ F_{1,2}, F_{1,2}^c \}$\hfil}} & 
\vbox to2.5ex{\vspace{1pt}\vfil\hbox to11ex{\hfil $ f_{1,2} $\hfil}} & 
\vbox to2.5ex{\vspace{1pt}\vfil\hbox to11ex{\hfil $ \tilde f_{1,2} $\hfil}} &  
\vbox to2.5ex{\vspace{1pt}\vfil\hbox to11ex{\hfil $ \tilde f^c_{1,2}  $\hfil}} & 
\vbox to2.5ex{\vspace{1pt}\vfil\hbox to11ex{\hfil $ f^c_{1,2} $ \hfil}} \\

\cline{1-5}
\end{tabular}
\caption{Bc's at $y=(0,\pi R)$ for 5D fields  with $\pm$ corresponding to Neumann/Dirichlet.
Only the $(+,+)$ fields have a 0-mode, and the KK mass spectrum ($n \geqslant 0 $) is:
$m_n=n/R$ for $(+,+)$ fields; $(2n+1)/2R$ for $(+,-)$ and $(-,+)$;  and $(n+1)/R$ for $(-,-)$.
$f_{1,2}$ stands for all 1st/2nd
generation fermions and $\tilde f_{1,2}$ their 4D $\mathcal{N}=1$ sfermion partners.
States in the last two columns correspond to the extra 5D SUSY partners. 
\label{tab:BoundaryConditions}}
\end{table}

On the other hand, the 3rd generation of matter is fully localized on the $y=0$ brane and therefore 3rd generation sfermions only pick up masses
at 1-loop from radiative corrections involving bulk fields.  Notice that due to the different localization of the 1st and 2nd generations compared to the 3rd, a natural hierarchy between sfermions is present in the theory.   Due to the $\mathcal{N}=2$ structure of the bulk, Yukawa interactions between Higgs and matter supermultiplets cannot be written as bulk terms but need to be localized on the $y=0$ brane -- the detailed structure of the Yukawa couplings in MNSUSY will be explored in Section~\ref{sec:yukawas}.

\begin{figure}
	\centering
	\includegraphics[width=5in]{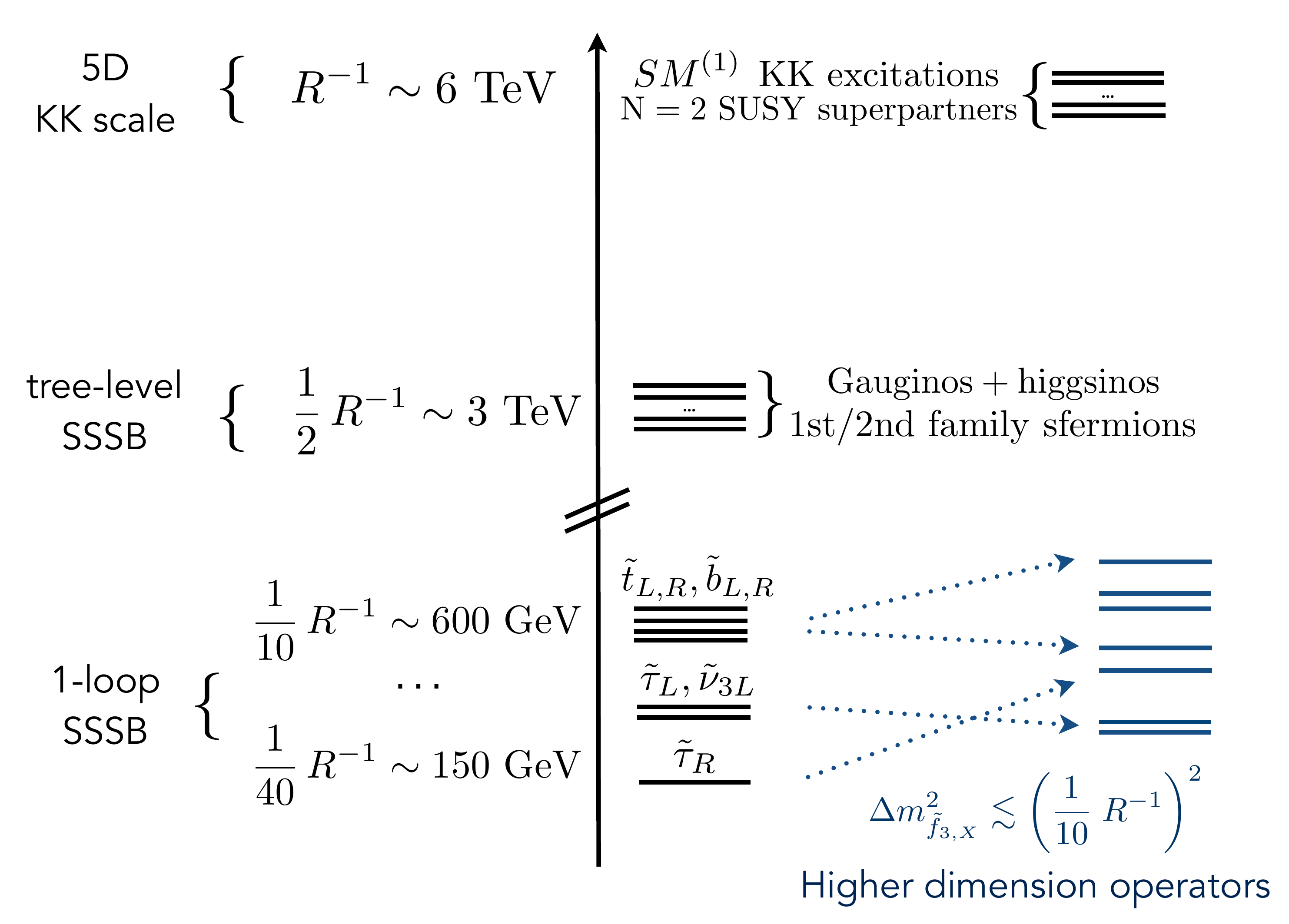}
\caption{The schematic spectrum of new states in the 5D SSSB model with an example KK scale of $1/R\sim 6~\TeV$. The MSSM-like gauginos and Higgsinos get tree-level SSSB Dirac masses at the scale $1/(2R)$ by pairing with their 5D conjugates. The lightest modes of the MSSM-like 1st and 2nd generation sfermions also appear at $1/(2R)$, along with their 5D SUSY conjugate scalar partners. The brane-localized 3rd generation sfermions get masses from SSSB at loop level, making the 3rd generation squarks about five times lighter than the gauginos. Although the SSSB tree-level and 1-loop contributions are fixed relative to $1/R$, HDOs can contribute to the 3rd generation sfermion masses at a similar order of magnitude and with undetermined coefficients, so that only the overall scale of the 3rd generation sfermion spectrum is predicted. At the KK scale $1/R$ the first SM KK excitations and additional 5D SUSY partners appear.} 
\label{fig:MinimalSpectrum} 
\end{figure}

The 1-loop contributions to the scalar masses of brane localized fields, as well as to the mass-squared parameters of the scalar components of $H_{u,d}$,
are given by similar expressions to Eq.~(\ref{eq:softmassg3}) but generalized to include extra gauge and Yukawa interactions:
\begin{equation}
	\delta{\tilde m_i}^2  \simeq  \frac{7 \zeta(3)}{16 \pi^4 R^2} \biggl( \sum_{I=1,2,3} C_I(i) g_I^2 + C_t(i) y_t^2 \biggr) ,
\label{eq:softmasssq}
\end{equation}
where $C(\bar U_3) = \{ 4/9 , 0 , 4/3 , 1\}$, $C(\bar D_3) = \{ 1/9 , 0 , 4/3 , 0 \}$, $C(\bar E_3) = \{ 1, 0, 0 , 0\}$,
$C(L_3) = \{ 1/4, 3/4, 0 , 0\}$, $C(Q_3) = \{ 1/36, 3/4, 4/3 , 1/2\}$ and for the 0-mode Higgs scalar components
$C(H_{u,d}) = \{ 1/4, 3/4, 0, 0 \}$ \cite{Antoniadis:1998sd,Delgado:1998qr}.
As mentioned in Section~\ref{sec:oneloopSSSB}, 3rd generation squarks receive the dominant part of their mass from bulk $SU(3)$ loops,
giving $\tilde m^2_{Q_3,U_3} \approx 1/(10R)^2$, with a splitting between $Q_3$ and $U_3$ due to the top Yukawa and $SU(2)$
contributions.
On the other hand, the right-handed stau gets the smallest 1-loop SSSB mass due to its SM gauge quantum numbers and tiny Yukawa interactions, 
\begin{equation}
	m^2_{\tilde \tau_R} \approx \left( \frac{1}{40 R} \right)^2 \approx \left( \frac{m_{ \tilde t}}{4} \right)^2 .
\end{equation}
However, while the pure SSSB limit is an interesting spectrum to study, couplings to $F_X$ of the form of Eq.~(\ref{eq:stopHDO}) can contribute to the masses of all of the third generation brane-localized sfermions. This sets a scale $\tilde m_{f_3}^2 \sim \Delta \tilde m^2_{U_3}$ (see Eq.~(\ref{eq:stopmassHDO}))
for the third generation sfermion masses with a non-predictive ordering of the spectrum.

Although in this work we will only consider the case that the 3rd generation is fully localized on the brane and the 1st and 2nd propagate in the bulk, small variations of this localization, motivated for instance by flavor constraints \cite{MNSUSYflavor}, are possible provided
the spectrum is \emph{locally} in the 5th dimension free of gravitational anomalies (rather than just globally free once integrated across the extra dimension) in order to avoid generating brane-localized FI terms \cite{Ghilencea:2001bw,Barbieri:2002ic,Marti:2002ar}.
As discussed in \cite{MNSUSYflavor}, some interesting possibilities consist of allowing part of the 3rd generation to propagate in the bulk,
or to brane-localize part or all the 1st and 2nd generations. 
Although the latter possibility does not correspond to a strictly `natural SUSY' spectrum, 
the limits on 1st and 2nd generation squarks can still be rather weak when the gluino is heavy,
and this model provides an attractive realisation of the `super-safe' Dirac gluino scenario \cite{Kribs:2012gx}.

Notice that we focus on the case where all states are either exactly localized on one of the branes or are allowed to propagate uniformly in the bulk.
However, when bulk masses are allowed for the bulk hypermultiplets, states can be quasi-localized toward the $y=0$ or $y=\pi R$ brane~\cite{Hebecker:2002re,Barbieri:2002sw}.  Forbidding these masses in our model is technically natural since the bulk parity $P_5$~\cite{Barbieri:2002sw}
(a symmetry that corresponds to a reflection about any point of the bulk in the limit $R \rightarrow \infty$)
is broken only globally by the inequivalence of the $y=0$ and $y=\pi R$ branes, but interesting properties arise in models where these mass terms are included and the 3rd generation is only quasi-localized.  For example, as studied in Refs.~\cite{Barbieri:2002sw,Barbieri:2003kn}, the propagation of quasi-localized 3rd generation sfermions into the bulk allows small tree-level SUSY breaking masses from the Scherk-Schwarz bc's in addition to the loop-level contributions from bulk multiplets.
Phenomenologically, such contributions play a very similar role to the $F_X$ shifts in the brane-localized masses we consider in this work, and most of our results apply straightforwardly to the quasi-localized models.




\subsection{Yukawa Couplings}
\label{sec:yukawas}

There are a variety of possibilities for realizing the Yukawa structure in a 5D SUSY set-up, and in MNSUSY some of the SM flavor structure can be explained by the localization of the third generation and the presence of $F_X$. We will focus in particular on the limit where $H_d$ does not obtain a vev and behaves as an inert doublet,
corresponding to an enhanced $U(1)_{H_d}$ (or $\mathbb{Z}_{k}$) symmetry of the theory. In this theory the down-like quarks will couple directly to $H_u$ through HDO, and the phenomenology differs substantially from the $\tan\beta \to \infty$ limit of the MSSM; in particular, rare flavor processes are \emph{not} enhanced by powers of $\tan\beta$ even though the Higgs sector itself
is realising the $\tan\beta \to \infty$ limit.

\subsubsection*{Up-Type Yukawas}

Yukawa interactions for up-like states cannot be written as bulk interactions, since they are forbidden by the extended bulk $\mathcal{N}=2$ SUSY, and must be generated instead as HDOs localized on the orbifold branes.
For example, Yukawa couplings for the up-like quark sector are given in the superpotential by (here $i,j=1,2$)
\begin{equation}
	W\  \supset\  \delta(y) \frac{\tilde{g} H_u}{\sqrt{M_*}}
	\left\{
	\tilde y_{33} Q_3 \bar U_3 + \tilde y_{3i} \sqrt{ \frac{3\pi/2}{M_*} } Q_3 \bar U_i +
	\tilde y_{i3} \sqrt{ \frac{3\pi/2}{M_*} } Q_i \bar U_3 + \tilde y_{ij} \frac{3\pi/2}{M_*} Q_i \bar U_j \right\} .
\label{eq:upyukawas}
\end{equation}
These bulk-brane interactions are volume suppressed, as reflected in the dimensionality of the couplings, and the 4D effective top Yukawa
naturally takes a value $y_t \sim \tilde{g}/\sqrt{M_* \pi R} \sim 1$, while those involving the 1st and 2nd generation have a further volume supression by a factor $\sim M_* R$.

Of course the relative values of the Yukawa couplings are \emph{not} fully explained by this geometrical arrangement of multiplets, but for example extensions to 6D orbifold models \cite{Hall:2001rz} (with one dimension having a SSSB twist), or the inclusion of traditional Froggatt-Nielsen style broken flavor symmetries \cite{Froggatt:1978nt} acting on the 1st and 2nd
generations can accommodate/explain the peculiarities of the  SM fermion
states and their mixings. 


\subsubsection*{Non-Holomorphic Down-Type Yukawas}
\label{sec:downyukawas}

In models where $H_d$ does not get a vev, the down-type quark and lepton masses and mixings cannot
arise from interactions with $H_d$.  We instead consider the case where down-type Yukawas are generated via non-holomorphic HDOs in the \kahler potential involving both $H_u$ and the SM-singlet $X$ \cite{Davies:2011mp}. For example, for down-type quarks 
\begin{multline}
	\mathcal{K}\ \supset\ \delta(y) \frac{\tilde{g}^2 X^\dagger H_u^\dagger}{\sqrt{3 \pi / 2} M_*^{5/2}}
	\left\{ \right.
	\hat y_{33} Q_3 \bar D_3 + \sqrt{\frac{3\pi/2}{M_*}} \hat y_{3i} Q_3 \bar D_i + \\
	\left. \sqrt{\frac{3\pi/2}{M_*}}\hat y_{i3} Q_i \bar D_3 + \frac{3\pi/2}{M_*}\hat y_{ij} Q_i \bar D_j \right\}
	+ {\rm h. c.} 
\label{eq:downyukawas}
\end{multline}
($i,j=1,2$) and similarly for leptons.
As for up-type Yukawa couplings, this structure implies a $\sim 1/(M_* R)$ suppression of the down-type couplings of the 1st and 2nd generation compared to that of the 3rd.  Moreover, the different structure of the up- and down-type Yukawas implies that the latter are naturally suppressed compared to the former. For example, the ratio of the 4D effective bottom and top couplings is naturally 
\begin{equation}
	\frac{y_b}{y_t} \approx \frac{f_X}{\sqrt{3\pi/2}(M_* \pi R)^{3/2}} \approx \frac{1}{300} \left(\frac{25}{M_* \pi R}\right )^{3/2}f_X .
\end{equation}
Reproducing the bottom quark mass may therefore require a slightly larger coupling $\hat{y}_{33} \sim 7$ than expected from the strong coupling expansion in $\tilde g$, (see Eq.~(\ref{eq:NDA})). If the extra source of strong coupling feeds in directly to all the other HDOs, then the appropriate strong coupling expansion is in terms of $\tilde g' \sim \tilde g \sqrt{\hat{y}_{33}}$. If it feeds in only through loops involving the operator in Eq.~(\ref{eq:downyukawas}), then the previous estimates are correct to $\mathcal{O}(1)$.
Both of these cases are viable and we can treat the former by allowing coefficients as large as $c\sim 7$ for the HDOs communicating soft masses from $F_X$ like in Eq.~(\ref{eq:stopHDO}).

This strong suppression of the down-like and lepton couplings relative to the up-like couplings is an interesting generic feature of our model, and we see that there are new opportunities for flavor model building in MNSUSY models, a large topic that is beyond the scope 
of this work. We also remark that although we do not utilise the possibility here, Yukawa couplings for the 1st and 2nd generations may also be written as brane-localized interactions on the $y=\pi R$ brane, making use of the different $\mathcal{N}=1'$ SUSY that is preserved there~\cite{Barbieri:2000vh}. If the third-generation is only quasi-localized, the same mechanism can also be used for the bottom Yukawa \cite{Barbieri:2000vh}, and these models have similar phenomenology to the case studied here.


\subsection{$U(1)_R$ Symmetry and Dirac Gauginos}
\label{sec:U(1)R}

The choice of bc's required for maximal SSSB is an enhanced symmetry point, preserving a $U(1)$ subgroup of the $SU(2)_R$ $R$-symmetry present in $\mathcal{N}=2$ SUSY.  The $R$-charges of the different fields are shown in Table~\ref{tab:Rcharges}.
\begin{table}[h]
  \begin{center}
    \begin{tabular}{ l | c | c }
	$\mathcal{N}=1$ superfield & Boson & Fermion \\
	\hline
	$V^a = (V^a_\mu, \lambda^a)$ & $0$ & $+1$ \\
	$\Sigma^a = (\bar \sigma^a, \bar \lambda^a)$ & $0$ & $-1$ \\
	$H_{u,d} = (h_{u,d}, \tilde h_{u,d})$ & $0$ & $-1$ \\
	$H^c_{u,d} = (h^c_{u,d}, \tilde h^c_{u,d})$ & $+2$ & $+1$ \\
	$F_{1,2,3} = (\tilde f_{1,2,3}, f_{1,2,3})$ & $+1$ & $0$ \\
	$F^c_{1,2} = (\tilde f^c_{1,2}, f^c_{1,2})$ & $+1$ & $0$ \\
	$X = (\phi_X, \psi_X)$ & $+2$ & $+1$ 
    \end{tabular}
    \caption{$R$-charges of relevant fields present in the theory (in $\mathcal{N}=1$ language).
	The pairs $(\lambda^a, \bar \lambda^a)$
	and $(\tilde h_{u,d}, \tilde h^c_{u,d})$ have opposite $R$-charges and partner resulting in
	Dirac gaugino and Higgsino masses. Note that $R(h_{u,d})=R(F_X)=0$.}
    \label{tab:Rcharges}  
  \end{center}
\end{table}

This $U(1)_R$ symmetry ensures that the two Weyl fermions present in an $\mathcal{N}=2$ vector supermultiplet ($\lambda^a$ and $\bar \lambda^a$)
and in Higgs hypermultiplets ($\tilde h_{u,d}$ and $\tilde h_{u,d}^c$) pair into Dirac fermions, giving
Dirac gauginos and Higgsinos with tree-level masses of size $1/(2R)$.
Moreover, $A$-terms are forbidden by the exact $R$-symmetry in maximal SSSB,
in contrast to models with a small $R$-symmetry twist, which generate large $A$-terms \cite{Murayama:2012jh}.
Notice that the remaining $R$-symmetry is not broken by the vev of $h_u$ (or of $h_d$ if it was non-zero),
by Yukawa interactions, or by the non-zero vev of the $F$-term of $X$. As discussed in Appendix~\ref{app:MaximalRadionMed}, radius stabilization and SUGRA interactions can preserve the $R$-symmetry, and we therefore focus primarily on the case that the $R$-symmetry is exact. If there are small breakings of the $R$-symmetry, as might result from non-perturbative violations of the global symmetries in a fundamental gravitational theory or gaugino condensation, their dominant effect can be parameterized in a shift $\delta \alpha \ll 1$ away from maximal twist. Small deviations from maximal twists can also arise in string compactifications where the twist is quantized in small units \cite{Kounnas:1988ye,Porrati:1989jk}. Nonvanishing $\delta\alpha$ will only be phenomenologically important when we study possible extensions with a chiral spectrum of fermions under the $R$-symmetry, leading to a state with mass controlled by $\delta \alpha$.

Another possibillity is that the radius stabilization sector explicitly breaks the $R$-symmetry. Then $A$-terms would be generated through couplings to $F_X$, and their sizes would be comparable to 3rd generation sfermion masses (see Eqs.(\ref{eq:stopHDO}) and (\ref{eq:stopmassHDO})).
Brane-localized gaugino Majorana masses would also be generated, but are a small perturbation on the large tree-level gluino Dirac mass $M_3 = 1/(2R)$ due to the brane-bulk 
overlap factor $1/(M_* R)$. In this work we focus on the case that the $R$-symmetry is preserved by the radius stabilization sector.

It is worth emphasising the contrast between Dirac gauginos in maximal SSSB and in supersoft 4D models that contain partial $\mathcal{N}=2$ SUSY \cite{Fox:2002bu}.  In 4D, the adjoint scalars present phenomenological difficulties: their imaginary components can be tachyonic unless protected from certain operators \cite{Alves:2015kia}
and the real components can remove the $D$-term contributions to the Higgs quartic unless gauginos obtain Majorana masses, thus breaking the $R$-symmetry \cite{Fox:2002bu,Nelson:2015cea}.
On the other hand, in 5D maximal SSSB both the real and imaginary components of the scalar are lifted without breaking the $R$-symmetry---the scalar adjoints obtain a mass through the SUSY  Stueckelberg mechanism with the KK gauge bosons \cite{Marti:2001iw}, which removes the tachyons, introduces pure Dirac masses for the gauginos, and preserves the $D$-term contributions to the Higgs quartic.

\subsection{The Scale of EWSB}
\label{sec:EWSB}

In the absence of SSSB loop contributions and HDOs, the Higgs scalar 0-modes are exactly massless
and their quartic interactions are given by the standard tree-level MSSM $D$-terms.
Radiative contributions are therefore crucial both to connect the scale of EWSB to the SSSB scale $1/R$
and to determine the viability of a $m_{h} \approx 125~\GeV$ Higgs. 

The structure of the corrections to the Higgs mass-squared parameter in comparison with the standard MSSM results
is most easily organized in the framework of matching the 5D theory to an effective 4D theory,
and this approach will make incorporating additional contributions to the Higgs potential from extended sectors
and other sources of SUSY breaking particularly simple.
The 5D SUSY theory valid at scales $\gtrsim 1/R$ 
can be matched to an effective 4D theory containing only the SM (including the Higgs 0-modes)
and 3rd generation superpartners at lower energies.
The leading contributions to the Higgs potential in this effective theory are similar to those of the MSSM,
and the properties of the 5D physics are all encapsulated in the matching.
As is well known, there are further important log-enhanced IR corrections to the Higgs potential that can be encapsulated
by integrating out the scalars at scale $m_{\tilde t}$ and running to match the measured SM couplings at the $Z$ and top poles.
In the case where only $H_u$ gets a vev, we find that there is a natural hierarchy between the EWSB and SSSB scales, with
$v \sim 1/(20 R)$.  However, the radiative potential does not favor EWSB and additional contributions from HDOs, as
discussed later, are required. 

To be concrete, our calculation of the $H_u$ zero-mode mass-squared parameter, $m_{H_u}^2$, in the effective theory includes a 1-loop EW contribution, $m^2_{H_u, EW}$,
and a two-loop Yukawa-mediated piece, $m^2_{H_u, y_t}$, of order $\sim y_t^4, g_3^2 y_t^2$ that also includes three-loop LL terms
$\sim (y_t^6, y_t^4 g_3^2, y_t^2 g_3^4) \times \log(x)^2$
where $\log(x)=\log\left(\frac{\tilde{m}_t}{1/\pi R}, \frac{m_t}{\tilde{m}_t}, \frac{m_t}{1/\pi R}\right)$ are treated as formally of the same order.
The 1-loop EW piece can be understood by integrating out the bulk Higgs and gauge KK modes at scale $1/(\pi R)$,
which generates the 1-loop positive mass-squared parameters in the effective theory given in Eq.~(\ref{eq:softmasssq}), and
for the Higgs is a positive contribution dominated by the $SU(2)$ sector,
\begin{equation}
m^2_{H_u, EW} = \frac{7 \zeta(3)}{16 \pi^4 R^2} \biggl( \frac{1}{4} g_1^2 + \frac{3}{4} g_2^2 \biggr)
			\approx \frac{7 \zeta(3)}{16 \pi^4 R^2} \frac{3}{4} g_2^2 \sim \left(\frac{1}{20 R}\right)^2
\label{eq:m2HuEW}
\end{equation}

Matching to the effective theory at 2-loops using the fixed order results of Ref.~\cite{Barbieri:2003kn} generates the non-logarithmically enhanced $(y_t^4, g_3^2 y_t^2)$ contribution to $m^2_{H_u, y_t}$. Running the soft masses down from $1/(\pi R)$ to the stop threshold, and running and matching the gauge couplings down through the stop threshold to the measured SM couplings at the top pole generates the remaining 2-loop and 3-loop LL contributions. As the log enhanced contributions are generated from running of the soft masses just as in the MSSM, it is not surprising that loops of stops generate a negative contribution to the Higgs mass of comparable size to that expected from a stop of similar mass in the MSSM with a low mediation scale.
The three-loop LL terms are an important contribution, giving a $\sim 50\%$ shift in the result compared to the fixed order calculation of Ref.~\cite{Barbieri:2003kn}, which can be understood as due to the significant running of $y_t$ and $g_3$ between $1/R$ and $m_t$ and the quartic dependence of the fixed-order result on these couplings. To within the theoretical error, we find the full result is numerically well approximated by evaluating the usual 1-loop MSSM formula at scale $\mu=1/(\pi R)$, using the 1-loop stop mass given by
Eq.~(\ref{eq:softmasssq}) and the  Yukawa and gauge couplings all evaluated at the $\overline{\rm DR}$ value given by SM running to the scale $\mu=1/(\pi R)$,
\beq
\label{eq:m2Huyt}
	m^2_{H_u, y_t} \approx \left. - \frac{3 y^2_{t,SM}(\mu)}{16\pi^2} \left[\tilde m^2_{Q_3}(\mu) + \tilde m^2_{U_3}(\mu)\right]\log\left[\frac{\mu^2}{\tilde m_{Q_3}(\mu) \tilde m_{U_3}(\mu)}\right]\right|_{\mu=1/\pi R} .
\eeq

\begin{figure}
\centering
\includegraphics[width=4in]{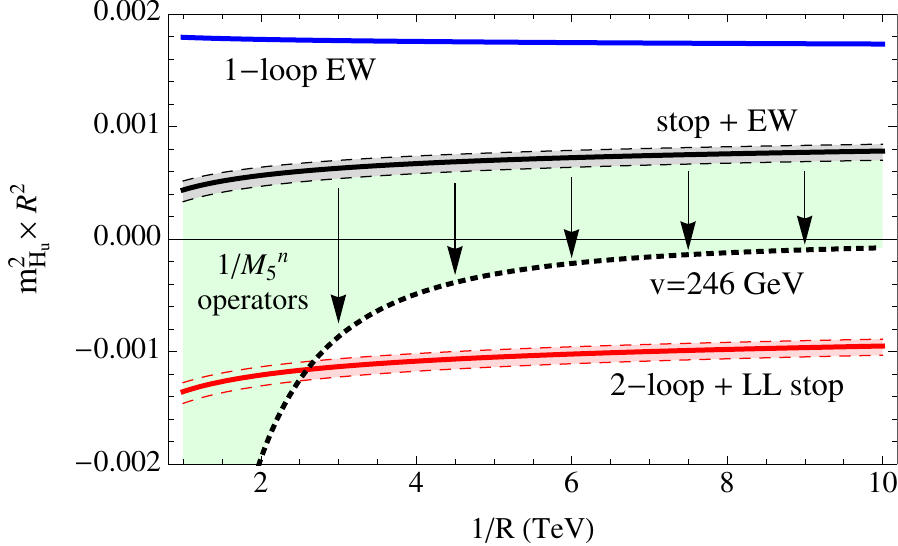}
\caption{
\label{fig:EWSBcond}
Contributions to the Higgs soft mass $m^2_{H_u}$ in units of $1/R^2$. The positive 1-loop EW contribution (blue) and the negative 2-loop + LL top-stop sector contribution (red) combine to give a positive, but tiny, mass squared (black), implying that the minimal model
is close to EWSB. Contributions from HDOs (see Eq.~(\ref{eq:m2HuHDO})) can lead to successful EWSB, indicated by the dotted black curve. The dashed bands show the uncertainty for $\bar{\rm MS}$ top mass $m_t(M_t) =160^{+5}_{-4} \gev$.} 
\end{figure}

The results of the full numerical evaluation of our calculation of EWSB are summarized in Figure~\ref{fig:EWSBcond}.
As shown in Figure~\ref{fig:EWSBcond}, these minimal contributions given by the pure SSSB contributions to $m^2_{H_u}$  do {\it not} lead to EWSB;  the positive 1-loop EW contribution
Eq.~(\ref{eq:m2HuEW}) is about twice the size of the negative radiative contribution from the Yukawa coupling
Eq.~(\ref{eq:m2Huyt}).

Nevertheless, the size of the resulting Higgs mass-squared parameter is tiny compared to the basic SUSY-breaking scale, $1/(\pi R)$, of the theory, with
$m^2_{H_u} R^2 \sim 5 \cdot 10^{-4}$, and thus the theory is very close to criticality with small perturbations to the basic picture being capable of
triggering EWSB with the correct vev.  Specifically we find 
an EWSB vacuum with the observed vev $\langle H_u \rangle = v/\sqrt{2}$ ($v \approx 246 \gev$) is obtained by taking into account HDOs
that couple the Higgs  and third generation squarks to SUSY breaking effects in the radius stabilisation sector, as described in Section~\ref{sec:Fx}.

The operator coupling $H_u$ directly to $F_X$ is of the form
\beq
\label{eq:m2HuHDO}
	{\cal K} \supset \delta(y) c_H \left(\frac{\tilde{g}^2}{M_*^3}\right) X^\dagger X H_u^\dagger H_u,
\eeq
and give a mass-squared of size
\begin{equation}
\Delta m^2_{H_u,X} \approx c_H f_X^2 \left(\frac{25}{M_* \pi R}\right)^2 \times \left(\frac{1}{80 R}\right)^2.
\end{equation}
This contribution can be of the same order as the radiative piece from the bulk EW sector given in Eq.~(\ref{eq:m2HuEW}).
Moreover, $F_X$ can also feed into the Higgs potential radiatively by shifting the stop masses as in Eq.~(\ref{eq:stopHDO}).
The shift in $\tilde m^2_{Q_3,U_3}$ due to $F_X$ is not volume suppressed and is enhanced by a factor of $\sim M_* R$ compared to $\Delta m^2_{H_u,X}$. The soft mass $\Delta m^2_{Q_3,U_3}$ feeds in to $m^2_{H_u}$ with a large 1-loop coefficient including a logarithmic sensitivity to the UV scale $M_*$ since the breaking of SUSY associated with $F_X$ does not preserve the non-local
nature of SSSB\footnote{Computing the radiative shift in the stop mass is equivalent to computing the mixing of the operators Eq.~(\ref{eq:m2HuHDO}) and Eq.~(\ref{eq:stopHDO}) running down from the scale $M_*$, which is a local 4D effect involving only states localized on the brane.},
\beq
\Delta m^2_{H_u, \tilde{t}} \approx -\frac{3 y_t^2}{16\pi^2}\log\left(\frac{M_*^2}{\tilde m_{Q_3} \tilde m_{U_3}}\right)(\Delta \tilde m^2_{Q_3} + \Delta \tilde m^2_{U_3}) \approx \frac{1}{10}(\Delta \tilde m^2_{Q_3} + \Delta \tilde m^2_{U_3}).
\label{eq:m2hRadiativeFx}
\eeq
This contribution is comparable to the direct mediation of  Eq.~(\ref{eq:m2HuEW}) and can on its own  drive EWSB when coupling to $F_X$ increases ${m}^2_{\tilde t}$ from the pure SSSB value of ${m}_{\tilde t} \sim 1/(10 R)$ to roughly $m_{\tilde t} \sim 1/(8 R)$.

Therefore we can use the $\mathcal O(1)$ freedom in the coefficients of the $F_X$ contributions to the squark masses to drive EWSB. The natural scale for the soft mass set by SSSB is $m^2_{H_u}\approx 1 / (20 R)^2$, and the correct value for EWSB is obtained by including the $F_X$ contributions to the Higgs potential. These two cancelling contributions are naturally of comparable size, and it is only when $1/R\gg 4~\TeV$ 
that the residual value of $m^2_{H_u}$ is too large and a tuning needs to be introduced to cancel the two contributions to give the observed weak scale. Phenomenologically, this approach has very similar effects to giving the 3rd generation squarks small tree-level SSSB masses by quasi-localization \cite{Barbieri:2002sw,Barbieri:2003kn}.

\subsection{The SM-like 125 GeV Higgs}
\label{sec:higgsMassBarbieri}

The radiative contributions to the Higgs pole mass are IR dominated, and at LL match exactly the MSSM radiative corrections. We take equal masses $\tilde m^2_{Q_3} \approx \tilde m^2_{U_3}$ for the stops, which holds to a good approximation in the pure SSSB model. The stop mixing parameters $X_t$ vanish because of the $R$-symmetry. The structure of the Higgs sector with only a $H_u$ vev means there are no Higgs mixing effects and the $y_b$-dependent corrections are negligible (unlike the MSSM at large $\tan\beta$). We define an effective Higgs quartic interaction by $m^2_{h} = V_{\rm eff}''(|H_u|^2) v^2 \equiv  2\hat\lambda v^2 $, with $\langle H_u \rangle = v/\sqrt{2}$ and the SUSY tree-level value $\hat\lambda_0 = \frac{g_1^2 + g_2^2}{8}$. We include the all-orders LL contribution to the Higgs mass below the scale $m^2_{\tilde t}$, which matches the MSSM result for single scale SUSY \cite{Carena:1995bx,Sasaki:1991qu} and includes the important leading EW correction,
\begin{equation}
	\delta \hat \lambda_{\rm EW} = -\frac{3 y_t^2}{64 \pi^2} (g^2 + g^{\prime 2}) \log\left[\frac{\hat m^2_{\tilde t}}{m_t^2}\right],
\end{equation}
which gives a $\sim5\%$ shift in the Higgs mass.

Above the scale $m^2_{\tilde t}$, there is sensitivity to physics up to the compactification scale $1/(\pi R)$ starting at two-loop order. We include the fixed two-loop-order LL contributions at $\mathcal{O}([g_3^2 y_t^4, y_t^6] \log(...)^2)$ extracted from the results of \cite{Barbieri:2003kn}, 
\beq
\label{eq:lambdaYukawa}
\delta \hat \lambda_{UV} = \frac{3 y_t^4}{(16\pi^2)}\left(-\frac{g_3^2}{\pi^2}\left(\log[(m_{\tilde t}\pi R)^2]^2+2\log[(m_{\tilde t}\pi R)^2]\log[m_{\tilde t}^2/m_t^2]\right)+\frac{y_t^2}{8\pi^2}\log[(m_{\tilde t}\pi R)^2]\right).
\eeq
This gives a $\sim 2\%$ correction to the mass. We find the fixed order one-loop LL $\mathcal{O}(g_{2,1}^4\log(...))$ electroweak corrections sensitive to $1/(\pi R)$ have a $\lesssim 1\%$ effect.

We can describe the origin of these terms in the language of matching to the effective 4D theory to illuminate the connection with MSSM results and to clarify how we will incorporate other corrections beyond the minimal SSSB spectrum. To obtain the two-loop LL result, the heavy KK modes can be integrated out at 1-loop at the scale $1/(\pi R)$, generating the 1-loop 3rd generation sfermion masses and Higgs soft mass in the effective theory; at this order the other thresholds can be ignored (thus at this level there is no sensitivity to 5D physics except through the loop generated stop masses). The one and two-loop LL terms are generated by running through and integrating out the stop thresholds then running to match the SM couplings, and clearly will have the same form as the MSSM with heavy Higgsinos and gauginos. When the stop masses are increased by HDOs involving $F_X$ as given by Eq.~(\ref{eq:stopmassHDO}), the effect on the radiatively generated quartic can be completely included at this order
by replacing $\tilde{m}^2_{Q_3,U_3} \rightarrow \tilde{m}^2_{Q_3,U_3~\rm(SSSB)} + \Delta m^2_{Q_3,U_3}$.

To estimate the uncertainty in our calculation, we use the experimental uncertainty in the top pole mass $m_t = 173.2 \pm 0.8~\GeV$ \cite{PDG} combined with an estimate of the theoretical uncertainty given by the magnitude of the leading MSSM next-to-LL contribution to the Higgs mass~\cite{Carena:1995bx}.

\begin{figure}
\centering
\includegraphics[width=4in]{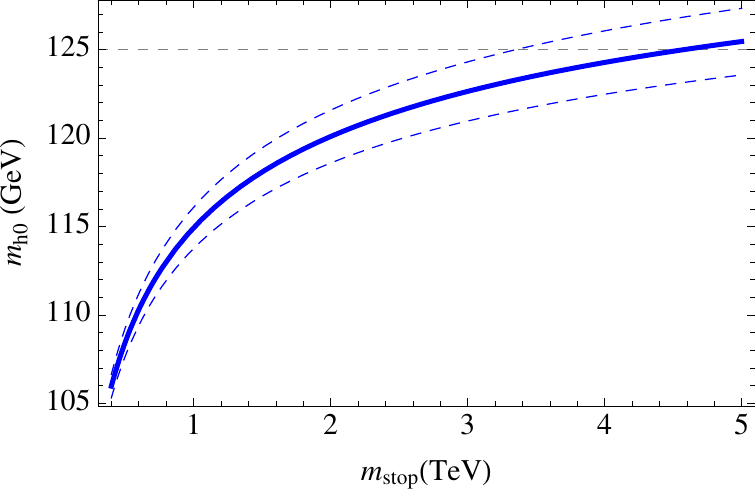}
\caption{\label{fig:HiggsPoleSSSB}Higgs pole mass in pure SSSB as described in Section~\ref{sec:higgsMassBarbieri}. The horizontal axis is the lightest stop soft mass. Here we have taken the $F_X$ contribution to the stop masses to vanish, so the stop mass is related to the compactification scale $1/R$ just by Eq.~(\ref{eq:softmasssq}) giving $m_{\tilde t} \approx 1/(10R)$. The dashed bands show the combined experimental and theoretical uncertainty. The experimentally observed value $m_{h^0} = 125 \gev$ is indicated.} 
\end{figure}

Thus we find that the Higgs mass prediction in the pure SSSB model, shown in Figure~\ref{fig:HiggsPoleSSSB}, matches well with the large $\tan\beta$ MSSM predictions. Note that while general models with Dirac gauginos can have additional enhancements over the MSSM contributions to the quartic \cite{Bertuzzo:2014bwa}, in MNSUSY there is not freedom to enhance the trilinear couplings or the adjoint masses to make these large.
 Unfortunately this implies that stops $\gtrsim 3~\TeV$ are necessary to obtain $m_{h} \approx 125~\GeV$, corresponding to a compactification scale $1/R \gtrsim 30~\TeV$. 
At such large stop masses the natural value for the Higgs vev is far above the measured weak scale, and the theory will be tuned to the few percent level. Although this is significantly less tuning than typical MSSM-like theories with comparable stop masses, we will be motivated to consider extensions that can increase the Higgs pole mass without requiring such large contributions from the stop sector.  The stops can also be lighter if large $A$-terms are generated, $A_t^2 \gtrsim m^2_{\tilde t}$, which can occur if the $X$ couplings break the $R$-symmetry. Because the leading radiative contributions to the Higgs pole mass are generated in the effective 4D theory containing the third generation squarks, the effects of including such $A$-terms can be incorporated using the appropriate 1-loop MSSM formula, and the range of viable parameters is similar to the MSSM. However, when large $A$-terms are generated by $F_X$, they feed into the Higgs soft mass with large logarithms sensitive to the fundamental scale $M_*$ and will dominate the tuning for $A_t^2 \gtrsim m^2_{\tilde t}$, removing the UV insensitivity that is one of the principle advantages of SSSB and leading to little improvement in the tuning. Instead, we focus in the following sections on extensions to the minimal field content that can increase the Higgs mass for light stops without substantially affecting the tuning or reintroducing further log sensitivity to $M_*$.

\section{Extended Sectors}
\label{sec:extensions}

The minimal version of MNSUSY allows for EWSB to occur with a low level of fine-tuning for a compactification scale $1/R \approx 4 - 10 \tev$, as can be clearly seen from Figure~\ref{fig:EWSBcond}. However, for the minimal field content, the contributions to the Higgs quartic coupling are too small to be consistent
with a $125 \gev$ Higgs in the region of low tuning. 

We are therefore consider extending the minimal version of MNSUSY to give an extra contribution to the Higgs quartic without affecting significantly the level of fine-tuning.
In the following, we describe three distinct possibilities for a natural theory of EWSB that
accomodates the experimentally measured Higgs mass:
(i) adding extra matter (in the form of vector-like leptons) coupling to the Higgs with $\mathcal{O}(1)$ Yukawa couplings, (ii) extending the gauge group of the theory with an extra $U(1)^\prime$ factor under which the Higgs is charged, and (iii) adding a brane-localized SM-singlet that couples to both Higgs doublets. Similar extensions have been studied in the context of realizing a $125~\GeV$ Higgs in 4D MSSM-like models \cite{Cheung:2012zq,Bertuzzo:2014sma, Hall:2011aa,Graham:2009gy, Martin:2009bg,Capdevilla:2015qwa,Bharucha:2013ela}, and we find that these mechanisms are easily implemented in MNSUSY, with potentially interesting consequences to the phenomenology discussed in Section~\ref{sec:pheno}.

\subsection{Vector-like Fermions Extension}
\label{sec:VL}

A simple mechanism to raise the physical Higgs mass $m_h$ consists in adding vector-like (VL) pairs of superfields with Yukawa couplings to the Higgs.
As noted in \cite{Martin:2009bg,Graham:2009gy}, a contribution to the Higgs quartic coupling, and therefore to $m_h$, will be radiatively generated with size
depending on the mass gap between the fermion and scalar components and the size of the new Yukawa.
We consider the simple case of adding two colorless $SU(2)_L$ doublets and singlets localized on one of the branes.
The extra field content, in $\mathcal{N}=1$ superfield notation, with their SM quantum numbers is the following:
\begin{equation}
\begin{aligned}
	{\tilde L} : \ & ({\bf 1}, {\bf 2}, -Y) \quad 			& {\tilde E} : \ & ({\bf 1}, {\bf 1}, Y+1/2)\\
	{\tilde L}^\prime : \ & ({\bf 1}, {\bf 2}, Y) \quad 	& {\tilde E}^\prime : \ & ({\bf 1}, {\bf 1}, -Y-1/2)~.
\end{aligned}
\end{equation}

With this field content, we can write a superpotential for the VL sector as follows\footnote{A Yukawa coupling involving ${\tilde L}$ and
${\tilde E}$ could also be written in the same way as lepton and down-type quark Yukawas. However, as we saw in Section~\ref{sec:yukawas}, their size
is parametrically suppressed compared to the up-type ones and therefore we neglect it here for simplicity.}:
\begin{equation}
	W \supset \delta(y) \left\{ \frac{\tilde k_u \tilde g}{M_*^{1/2}} H_u {\tilde L}^\prime {\tilde E}^\prime - \mu_L {\tilde L}^\prime {\tilde L} - \mu_E {\tilde E}^\prime {\tilde E} \right\}~,
\label{eq:WVL}
\end{equation}
where we have chosen to localise the new states on the $y=0$ brane (most of the following discussion applies also to localization in the $y=\pi R$ brane), and the 4D effective Yukawa coupling $k_u$ is given in terms of the fundamental parameters
as $k_u = \tilde k_u \tilde g (M_* \pi R)^{-1/2} \approx \tilde{k}_u$.

The new field content consists of two Dirac fermions with electric charge $\pm(Y+1/2)$ that couple to the Higgs
and one Dirac fermion with charge $\pm(Y-1/2)$ that does not,
together with their corresponding scalar partners.
We will concentrate on the case $Y=1/2$, which means that the mass eigenstates are
two fermions with charge $\pm 1$ ($\tau^\prime_1$, $\tau^\prime_2$)
and one neutral fermion ($\nu^\prime$) whose tree-level mass is equal to $\mu_L$. For $\mathcal{O}(1)$ values of the 4D effective coupling $k_u$, the
masses of the new fermions are such that $m_{\tau_1^\prime} < m_{\nu^\prime} \approx \mu_L < m_{\tau_2^\prime}$.
For simplicity, we impose an extra $\z$ symmetry on these new states to avoid their mixing with
SM leptons\footnote{The $\z$ symmetry will be slightly broken to allow the new states to decay.
See Section~\ref{sec:phenoVL} for details.}.
We refer to this version of the model as the \emph{VL-lepton} scenario.

A simple way to dynamically generate the VL masses $\mu_L$ and $\mu_E$ is to introduce a SM singlet in the 5D bulk, $K$, that couples to the new
VL states as follows:
\begin{equation}
	W \supset \delta(y) \frac{{\tilde \lambda}_K \tilde g}{M_*^{1/2}} K ({\tilde L}^\prime {\tilde L} + {\tilde E}^\prime {\tilde E})~.
\label{eq:W_SVL}
\end{equation}
At 1-loop, this interaction contributes to the soft masses of the scalar partners of the VL states.
Including all 1-loop contributions, these are given by the usual expression:
\begin{equation}
	\delta{\tilde m_i}^2 \simeq  \frac{7 \zeta(3)}{16 \pi^4 R^2} \left( \sum_{I=1,2,3} C_I(i) g_I^2 + C_{k_u}(i) k_u^2 + C_{\lambda_K}(i) \lambda_K^2 \right)
\label{eq:1loopsoftmass}
\end{equation}
with $C({\tilde L}) = \{ 1/4, 3/4, 0, 0, 1/2 \}$, $C({\tilde L}^\prime) = \{ 1/4, 3/4, 0, 1/2, 1/2 \}$,
$C({\tilde E}) = \{ 1, 0, 0, 0, 1/2 \}$, $C({\tilde E}^\prime) = \{ 1, 0, 0, 1, 1/2 \}$ \cite{Antoniadis:1998sd,Delgado:1998qr} and
where $\lambda_K = {\tilde \lambda}_K \tilde g (M_* \pi R)^{-1/2} \approx {\tilde \lambda}_K$ is the 4D effective coupling.
In turn, at two-loop order, the VL states generate a scalar potential for the $K$ field, in
much the same way as the brane localized top sector does for the Higgs,
with a negative soft mass and an $\mathcal{O}(1)$ quartic coupling.
This results in the 4D-normalized scalar component of $K$ getting a vev $\langle k \rangle \sim 10^2 \ \GeV$,
which leads to VL masses arising from Eq.~(\ref{eq:W_SVL}) as
\begin{equation}
	\mu_L = \mu_E = \frac{{\tilde \lambda}_K \tilde g}{\sqrt{M_* \pi R}} \langle k \rangle = \lambda_K \langle k \rangle \equiv \mu_{VL}~.
\end{equation}

\begin{figure}[h!]
	\centering
	\includegraphics[scale=0.35]{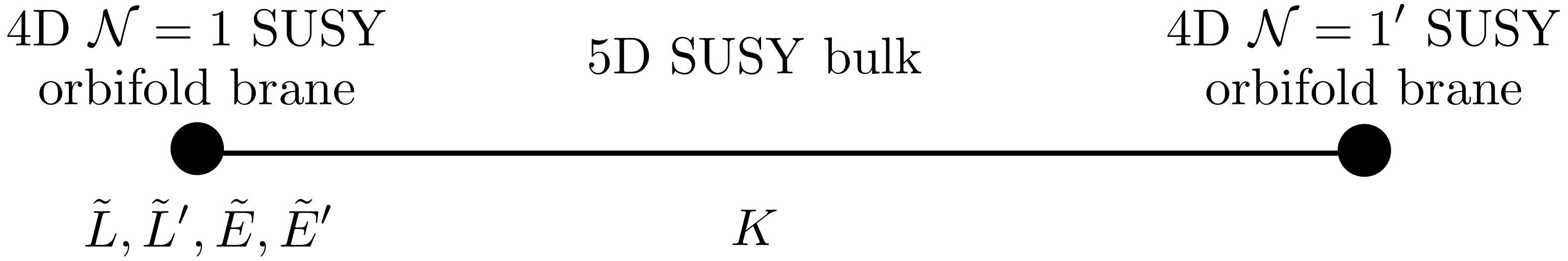}
\caption{Extra field content needed to implement the VL lepton variation:  A 5D bulk SM singlet hypermultiplet
	$K$ and a pair of vector-like $SU(2)_L$ doublets and singlets on the $y=0$ brane.} 
\label{fig:MinimalGeography_VL} 
\end{figure}

We take $\lambda_K \approx 2.0$ at the scale of the VL masses, which is consistent with the strong coupling expansion Eq.(\ref{eq:NDA}) at scale $M_*$. From now on, we also fix $\mu_{VL} = 350 \ \GeV$ for illustration, a natural value given the size of $\lambda_K$ and $\langle k \rangle$.
Figure \ref{fig:MinimalGeography_VL} shows the extra field content and its location.
Notice that this extension of the model that includes a bulk SM-singlet $K$ coupling to the VL fermions as specified in Eq.~(\ref{eq:W_SVL})
is particularly appealing because it generates both VL masses by the above mechanism and, at 1-loop, SUSY breaking masses for the scalar components.
This would also allow us to localise the VL states on a different brane from the SM-singlet $X$ (i.e. at $y=\pi R$), avoiding SUSY breaking contributions to the scalar masses
from $F_X$ that would result in the Higgs mass-squared parameter being UV sensitive (see Section~\ref{sec:EWSB}).
The statements made in this Section are largely independent of which brane the VL leptons are confined to, and we will assume
that the contribution to the scalar masses from $F_X$ is negligible compared to the contribution obtained radiatively from couplings to the bulk SM-singlet $K$.

In order to compute the extra contribution from the VL sector to the Higgs mass we use the 1-loop effective potential formalism in the 4D low energy theory.
The result depends on the size of the gap between fermion and scalar masses, i.e. on $1/R$, and crucially on the
size of the new Yukawa coupling, as $\delta m_h^2 \propto k_u^4$.  We add this contribution from the VL sector to the
one from the top-stop sector discussed in Section~\ref{sec:higgsMassBarbieri}.

\begin{figure}[h!]
	\centering
	\includegraphics[scale=0.6]{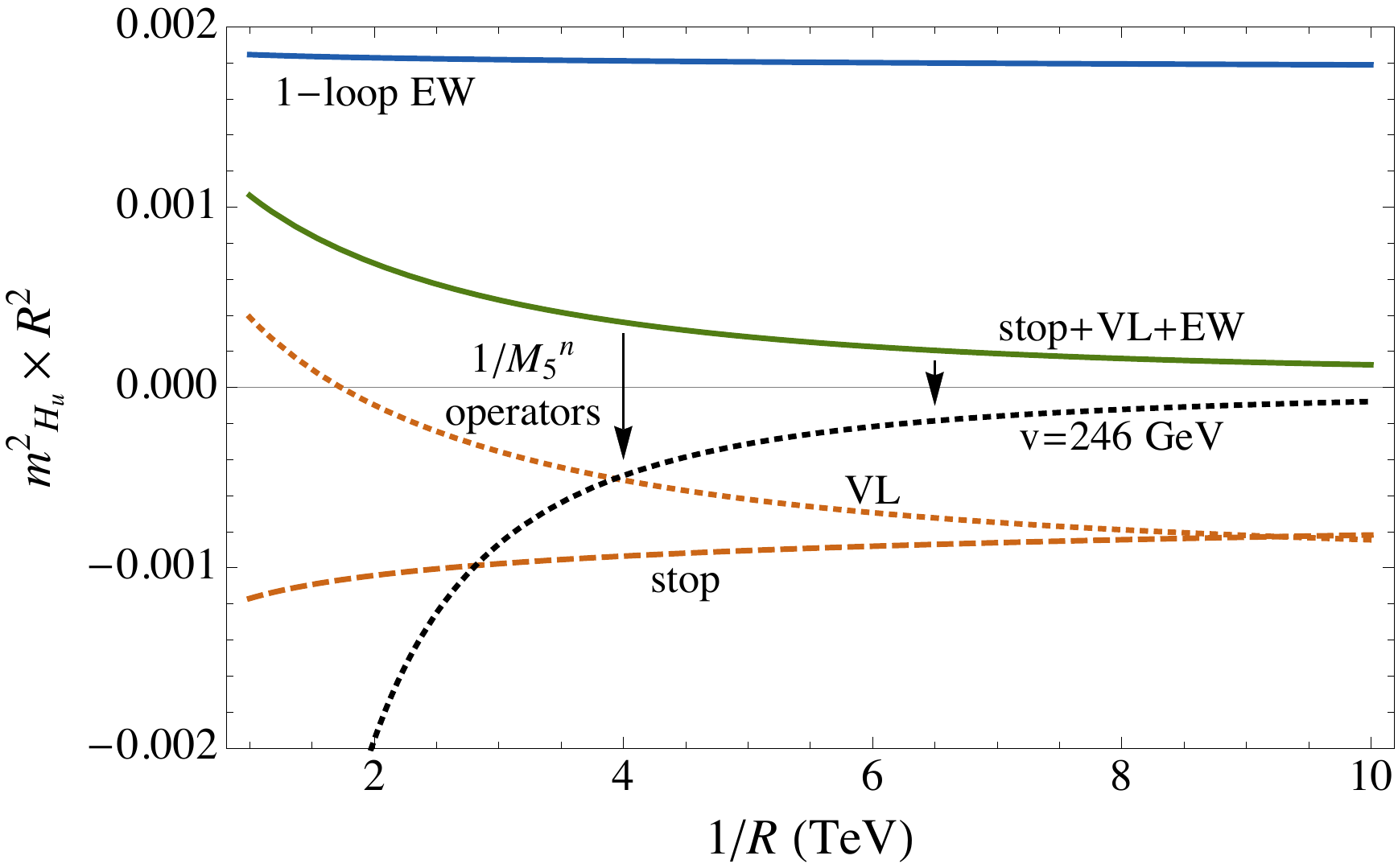}
\caption{Contributions to the Higgs soft mass squared, $m_H^2$, normalized to $1/R^2$ as a function of $1/R$.
	The blue line represents the 1-loop contribution from the EW sector, the dashed orange line the contribution from the top-stop sector
	and the dotted orange line that from the VL sector with the coupling $k_u$ chosen to give $m_h=125\GeV$. The green line is the sum of all three together and the dotted black line represents
	the correct value of the Higgs soft mass for successful EWSB (as achieved after adding the contributions from HDOs).} 
\label{fig:EWSB_withVLsector} 
\end{figure}

The new VL states also contribute to the Higgs soft mass parameter. Figure \ref{fig:EWSB_withVLsector} shows the different contributions to the soft mass-squared of the Higgs; the new sector gives a negative contribution to $m_H^2$ of approximately the same size
as that from the top-stop sector, leading to a total contribution closer to the true EWSB value compared to the previous version of the model
(see Figure~\ref{fig:EWSBcond}).
\begin{figure}[h!]
	\centering
	\includegraphics[scale=0.6]{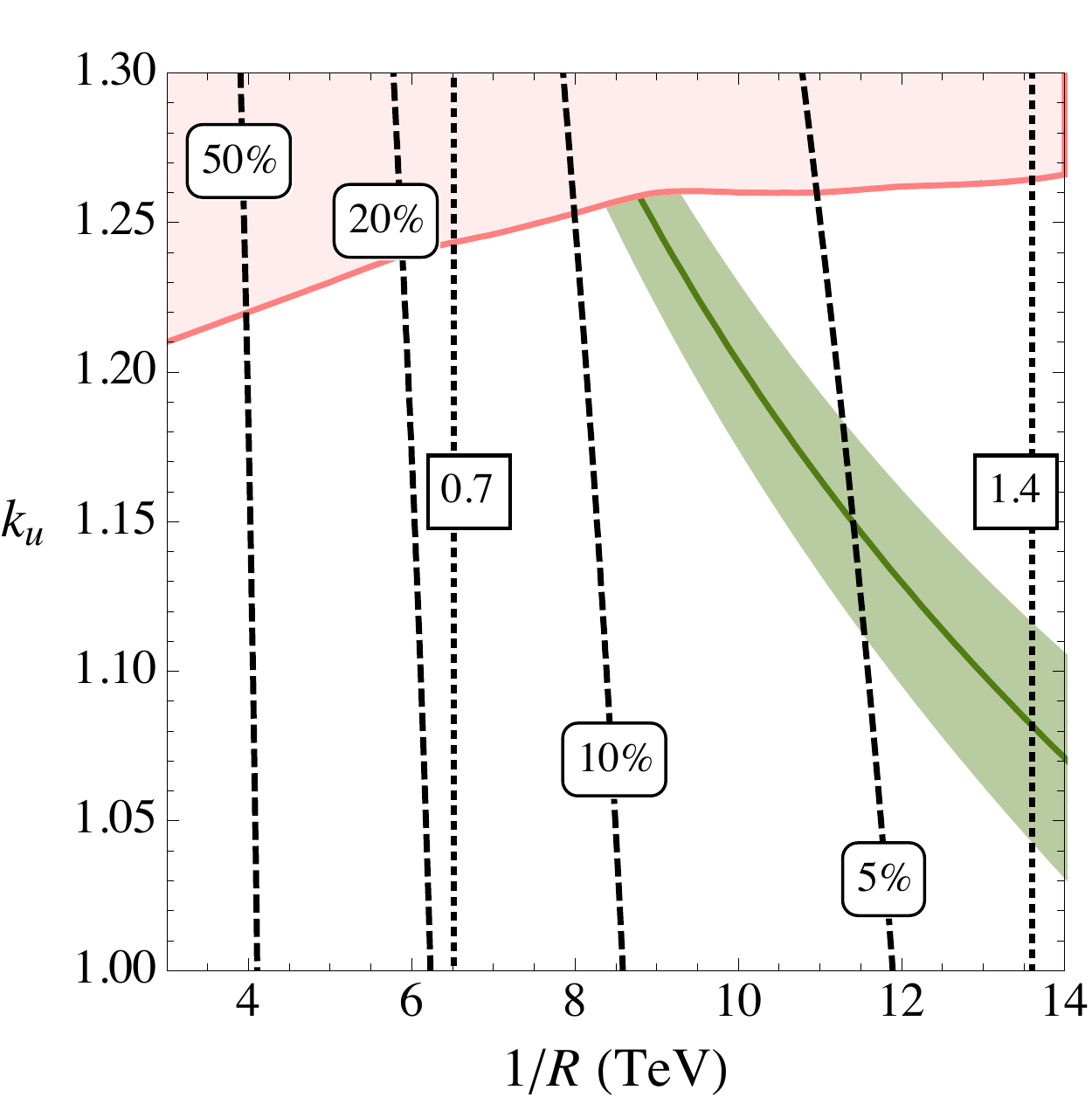}
\caption{The green area represents the region of parameter space where a 125 GeV mass
	for the Higgs is predicted, within the combined theoretical and experimental (mainly on $y_t$) uncertainty.
	The pink area is excluded by precision electroweak constraints, forbidding the region with too large $k_u$.
	Dashed lines represent contours of given tuning, as specified in the labels, and dotted lines denote the regions where stop masses
	are 0.7~TeV (around the current LHC limit) and 1.4~TeV (approximately the maximum stop mass to be probed by LHC14).} 
\label{fig:higssMassVL} 
\end{figure}
We adopt a conservative and measure of fine tuning, given by\footnote{This definition of the tuning measure corresponds to the usual Barbieri-Giudice measure \cite{Barbieri:1987fn} with $1/R^2$ and the soft mass scale of the new vector-like states $\tilde m^2_{VL}$ taken as the independent UV parameters.}:
\begin{equation}
	\Delta = \left\{ \left( \frac{ {m_{H, {\rm EW}}^2} + {m_{H, {\rm top}}^2}}{m_{H, {\rm exp}}^2} \right)^2 +
		\left( \frac{{m_{H, {\rm VL}}^2}}{m_{H, {\rm exp}}^2} \right)^2 \right\}^{-1/2}
\label{eq:tuningVL}
\end{equation}
where $m_{H, {\rm top}}^2$ is the contribution to the Higgs soft mass from the top sector, $m_{H, {\rm VL}}^2$ the
contribution from the VL sector and $m_{H, {\rm exp}}^2 \approx -(125 \ \GeV)^2/2$ the experimentally measured value.

Finally, Figure~\ref{fig:higssMassVL} shows the region of parameter space where one can achieve a 125 GeV Higgs
together with the level of fine tuning, quantified as specified in Eq.~(\ref{eq:tuningVL}).
As one can see, a model with the correct value of the Higgs mass and $\sim 10 \%$ tuning is possible in the
VL-lepton variation.

\subsection{$U(1)'$ Extension}
\label{sec:DTerms}

If the Higgs is charged under an additional gauge sector, the $D$-term generates additional contributions to the Higgs quartic that decouple as $\sim m^2_{\rm soft}/f^2$, where $f$ is the scale of the gauge group breaking and $m^2_{\rm soft}$ is the SUSY breaking mass coupling to the multiplet breaking the gauge group \cite{Batra:2003nj,Maloney:2004rc,Cheung:2012zq}. We will focus on the simple case of a new bulk $U(1)'$ gauge group with 5D gauge coupling $g_{X,5} \equiv g_{X,4}\sqrt{\pi R}$.  Collider and precision observables constrain the $Z'$ mass to be $m_{Z^\prime} \gtrsim 3 \tev$ \cite{ATLAS-CONF-2013-017,Chatrchyan:2012oaa}, which means SUSY breaking must be felt at scales $\sim 1/R$ for a sizeable non-decoupling effect.  A simple model where this occurs has the $U(1)'$ breaking driven by on-brane dynamics for bulk hypermultiplets $\Phi_1, \Phi_2$ that feel tree-level SSSB and have charge $Q_\Phi=\pm 1$ under the $U(1)'$. To be concrete we identify the $U(1)'$ with $T_{3R}$ normalized to $Q_{H_{u,d}}=\pm 1/2$, introducing sterile neutrinos to cancel anomalies.  We find that the Higgs pole mass can be lifted to $125~\GeV$ with $g_{X,4}\approx g_2$ and $m_{Z^\prime} \sim 1/R$ without substantially increasing the tuning of the weak scale.

In detail, a vev for $\phi_{1,2}$ (the scalar components of $\Phi_{1,2}$) can be induced by an interaction with a brane-localized singlet $V$, 
\beq
	W \supset  \frac{\lambda_V}{M_*} V \left(\Phi_1 \Phi_2 - \frac{f^2}{\pi R} \right) \delta(y)~.
\eeq
In the strong coupling expansion Eq.(\ref{eq:NDA}), $\lambda_V \sim \tilde{g}\sqrt{3\pi/2}$.
If the scalar components of $\Phi_{1,2}$ did not have SUSY-breaking bc's, this potential would introduce a SUSY preserving vev for $\phi_{1,2}$ and the $D$-term would decouple. In the presence of the SUSY breaking bc's, the boundary action can induce a SUSY breaking background and the $D$-term does not decouple. Some intuition for the behavior of $\phi_{1,2}$ can be obtained by truncating to the lightest scalar KK modes. The lightest modes have a SSSB mass $\tilde{m}^2_{\phi^{(0)}_{1,2}} = 1/(2R)^2$ and a potential generated from $F_V$,
\beq
	\mathcal{L}_4 \supset \left(\frac{\lambda_V}{M_* \pi R}\right)^2\left|\phi^{(0)}_1 \phi^{(0)}_2  - f^2 \right|^2~. 
\eeq
For $\lambda_V f^2/(M_* \pi R) \gtrsim 1/(2R)^2$, a vev in the $D$-flat direction $\phi_1^{(0)}= \phi_2^{(0)}\sim f$ will be generated. However, in this regime the brane-perturbation is strong and a truncated treatment of the lightest KK modes is no longer justified. 
Instead a full 5D calculation gives a kinked profile 
\beq
\phi_1(y) = \phi_2(y) = \phi_0 \frac{y-\pi R}{\pi R},\;\;\; \phi^2_0 = \left(f^2 - \frac{2 M^2_*}{\lambda_V^2}\right)\frac{1}{\pi R}
\eeq  
when $\phi_0^2 > 0$.  This results in an $F$-term for the singlet $F_V = M_*/(\lambda_V \pi R)$ and an $F$-term for the conjugate bulk fields proportional to the gradient $F_{\phi^c_{1,2}}= \phi_0/ (\pi R).$   The surviving $D$-term for the Higgs 0-mode can be determined by integrating out the tree-level fluctuations of $\phi_1$ and $\phi_2$ on this $y$-dependent background.  Defining a dimensionless $\omega = g_{X,5} \phi_0 \pi R= g_{X,4} \phi_0 (\pi R)^{3/2}$ this contribution to the Higgs quartic has the form 
\beq
\Delta \hat \lambda = \frac{g_{X,4}^2}{8}\left(\frac{4}{9} - \frac{8}{2835}\omega^2 + ... \right).
\eeq
An interesting feature of this model is that because the SUSY-breaking background $F_{\phi^c_{1,2}}$ never parametrically exceeds the vev $\phi(y)$ breaking the gauge group,  there is no regime where the $D$-term is completely re-coupled.

The parameter $\omega$ can be related directly to the mass of the lightest mode of the $Z'$ gauge boson in this background. For $\omega \ll 1$, the mass approaches the 4D result, $m^2_{Z'} = (4/3) \omega^2 (\pi R)^{-2}$, while for $\omega \gg 1$ the lightest $Z'$ state becomes localized away from the $y=0$ brane and asymptotically has mass $m^2_{Z'} \rightarrow \omega (\pi R)^{-2}.$ In the parameter range of interest $\omega\sim1$ and we evaluate $\Delta \hat \lambda$ and $m_{Z'}$ numerically from the full 5D equations of motion.

In addition to the tree-level contribution to the Higgs pole mass from the non-decoupling $D$-terms, there will be new loop-level contributions to the
Higgs soft mass from couplings to the $U(1)'$ gauge sector. If the gauge group were not broken, the contribution would have the same form as the
SSSB loop contribution from the SM gauge groups,  $\Delta m^2_{H,~U(1)'}= 7\zeta(3) g_{X,4}^2 / (64\pi^4 R^2)$.
When the gauge group is broken, the gauge states become localized toward the $y=\pi R$ brane, and the pure SSSB contribution is partially screened.
However, new sources of radiative SUSY breaking are introduced with $\phi_{1,2}$ acting as messengers to communicate $F_V$ and $F_{\phi^c_{1,2}}$ to the Higgs sector.
These contributions are cut-off at the scale of the $\phi_{1,2}$ masses $\sim \phi_0 \sqrt{\pi R} \sim 1/R$ and do not introduce any logarithmic sensitivity to $M_*$.
We evaluate the 5D propagators numerically in the $\phi(y)$ background to obtain the loop contributions to the Higgs mass, and find, in the parameter range of interest,
\begin{equation}
\Delta m^2_{H, U(1)'}\simeq 10^{-3} \times g_{X,4}^2 m_{Z'}^2~.
\end{equation}

\begin{figure}
\centering
 \includegraphics{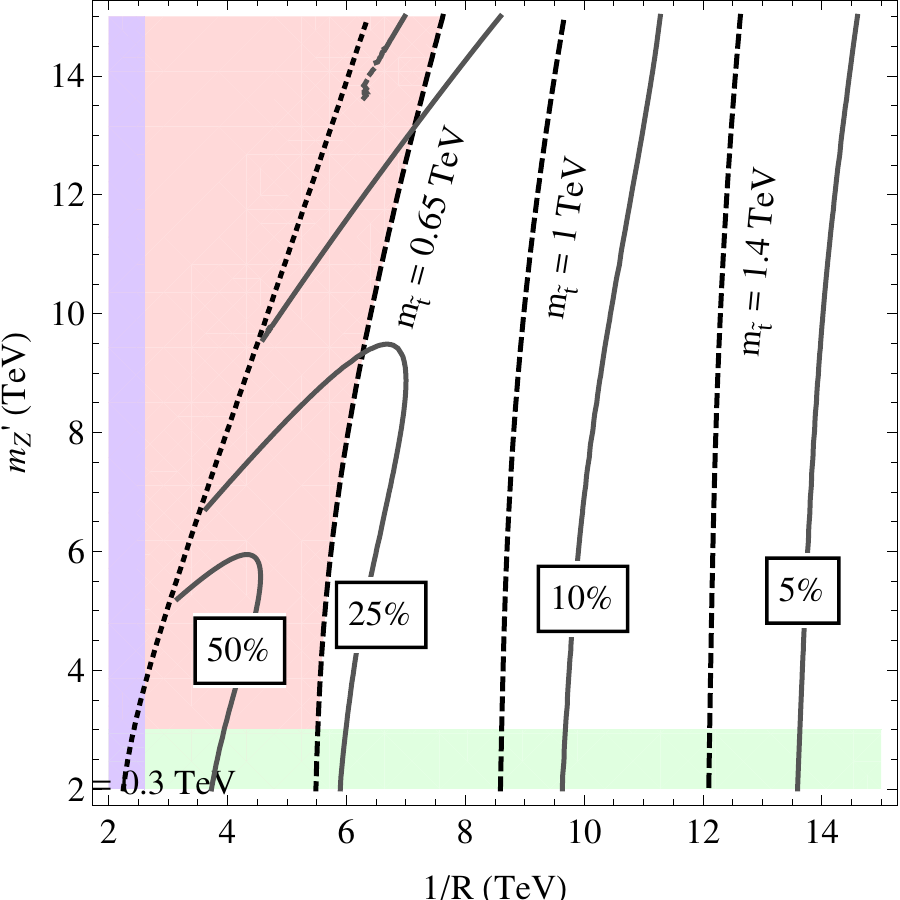}
\caption{\label{fig:tuningDTerms} Fine-tuning $\Delta^{-1}$ (solid lines) as a function of $1/R$ and the $Z'$ mass,  Eq.~(\ref{tuningeq}).   Iso-contours of stop mass are dashed. Limits from LHC8 searches for $\tilde{t}\to t+{\rm MET}$ \cite{Aad:2013ija,Chatrchyan:2013xna} (red) and $Z'$ resonance searches \cite{ATLAS-CONF-2013-017,Chatrchyan:2012oaa} (green) are shaded. Subdominant limits $m_{\tilde{g}}\approx 1/(2R)\gtrsim 1.3 \tev$ from $\tilde{g}\to t \overline{t}/b\overline{b}+{\rm MET}$ searches (blue) are also shaded \cite{ATLAS-CONF-2013-061,Chatrchyan:2013iqa}.}
\label{fig:tuningDterms}
\end{figure}

We evaluate the tuning of this model with respect to the shifts in the stop mass through HDOs and shifts in the
parameters of the $U(1)'$ sector as\footnote{This definition of the tuning measure corresponds to the usual Barbieri-Giudice measure \cite{Barbieri:1987fn} with $1/R^2$ and $m_{Z'}^2$ taken as the independent UV parameters.}
\beq
	\Delta = \left\{ \left( \frac{ {m_{H, {\rm EW}}^2} + {m_{H, {\rm top}}^2}}{m_{H, {\rm exp}}^2} \right)^2 +
		\left( \frac{{m_{H, U(1)'}^2}}{m_{H, {\rm exp}}^2} \right)^2 \right\}^{-1/2}
\label{tuningeq}
\eeq
The result is shown in Figure~\ref{fig:tuningDterms}. The tuning is driven by limits on direct production of stops, and the model is tuned at a level of $\sim 25\%$ (for rough LHC8 limits of $\sim 700~\GeV$ stops and $\sim 3~\TeV$ gluinos).

So far we have discussed how the extended gauge sector can lift the Higgs pole mass to the observed value. In fact, in the model described the extra states can play all the roles of the extra source of SUSY breaking so far parameterised by $F_X$. When $\lambda_V$ has its strong coupling value $\lambda_V \sim \tilde{g}\sqrt{3\pi/2}$, $F_V \sim~1/(\pi R)^2$ is a brane-localized $F$-term that can communicate soft masses through additional HDOs in addition to the predictive loop level IR  communication we have discussed. The tree-level potential due to $F_{\phi^c_{1,2}}$ and $F_V$ will generate a contribution to the radion potential of the Goldberger-Wise form \cite{Goldberger:1999uk}, which can allow the radion to be stabilised at vanishing CC. A hierarchy between $f$ and $M_*$ is technically natural, and the role of the tree-level potential in stabilising the radius may allow the relationship between scales $f\sim 1/R \ll M_*$ to be dynamically realised.  Thus we see that the $U(1)'$ extension is both minimally tuned and possesses attractive features from a theoretical perspective.

\subsection{Singlet Extension}
\label{sec:singlet}

It is well known that an additional tree-level contribution to the Higgs quartic is generated when the MSSM is extended to include a SM singlet coupling to the Higgs sector through a superpotential $\sim S H_u H_d$. This mechanism requires both $H_u$ and $H_d$ to obtain a vev, introducing complications beyond the models studied so far where only $H_u$ is responsible for all of the EWSB and Yukawa interactions. For completeness, we briefly study this mechanism in the context of MNSUSY theories.


To avoid decoupling the extra contributions to the Higgs mass, the singlet sector must feel SUSY breaking. Surprisingly, the simple possibility of introducing a bulk singlet with the scalars obtaining tree level SUSY breaking masses $\sim 1/2R$ does not generate a non-decoupling quartic\footnote{In particular, integrating out the scalar component of $S^c$ is crucial for the $\lambda_S^2$ contribution to the Higgs quartic to vanish. Notice that an analogous case does not happen in normal 4D models where only a single chiral multiplet $S$ is added without further 5D SUSY partners.}. Instead we consider the case where the chiral multiplet $S$ is localized on one of the branes, with superpotential
\begin{equation}
	W \supset \delta(y) \frac{{\tilde \lambda}_S \tilde g \sqrt{3\pi/2}}{M_*} S H_u H_d~,
\label{eq:singlet}
\end{equation}
which gives a 4D coupling for the zero modes $\lambda_S \approx 0.4 \tilde{\lambda}_S \left(\frac{25}{\pi M_* R}\right)^{1/2}$. Both the scalar and fermion components are massless at tree-level, but at 1-loop the scalar gets a SSSB mass due to its interaction with
the bulk Higgs doublets ${\tilde m}^2_S \approx 7 \zeta (3) \lambda_S^2 / (16 \pi^4 R^2)$ \cite{Antoniadis:1998sd,Delgado:1998qr} (we discuss the mass of the fermion in Section~\ref{sec:ChiralSinglets}). The additional contribution to the physical Higgs mass is
\begin{equation}
	\delta m_h^2 = \frac{v^2}{2} \lambda_S^2 \sin (2 \beta)^2~.
\label{eq:mhsinglet}
\end{equation}

A $125 \ \GeV$ physical Higgs mass is then achievable for moderately light stop masses. 
For example, when the compactification scale is $1/R \approx 4 \ \TeV$, a coupling $\lambda_S \approx 0.7$ and a value of $\tan \beta \approx 2$ 
lifts the Higgs to its observed mass.

This lifting, however, has a consequence for EWSB and tuning, as both soft masses $m_{H_u}^2$ and $m_{H_d}^2$ receive a
positive 1-loop contribution proportional to $\lambda_S^2$,
on top of the EW and top sector contributions previously discussed. 
This contribution is positive, equal for both $H_u$ and $H_d$, and comparable in size to the EW contribution,
\begin{equation}
	m_{H, S}^2 = \frac{2 \lambda_S^2}{3 g^2} m_{H, {\rm EW}}^2 = 0.8 ~m_{H, {\rm EW}}^2 \left(\frac{\lambda_S}{0.7}\right)^2.
\end{equation}
Despite this additional contribution disfavoring EWSB, we find that both Higgs doublets can achieve a non-zero vev due to negative SUSY breaking masses from HDOs of the form Eq.~(\ref{eq:m2HuHDO}).

In order to give a rough estimate of the tuning of this version of the model, we take the decoupling limit, such that only a light
Higgs boson remains in the low energy theory. Note that unlike 4D NMSSM models, we do not require $S$ to obtain a vev, as the Higgsino masses are obtained already through the bulk bc's. The condition for successful EWSB can be written as
\begin{equation}
	m_{H_u}^2 \sin^2 \beta + m_{H_d}^2 \cos^2 \beta = m_{H, {\rm exp}}^2~.
\end{equation}
In the spirit of the previous section, we then estimate the tuning as\footnote{This definition of the tuning measure corresponds to the usual Barbieri-Giudice measure \cite{Barbieri:1987fn} with $1/R^2$ and the soft mass of the singlet $\tilde{m}^2_{S}$ taken as the independent UV parameters.}
\begin{equation}
	\Delta \simeq \left\{ \left( \frac{m_{H, {\rm EW}}^2 + \sin^2 \beta \ m_{H, {\rm top}}^2}{m_{H, {\rm exp}}^2} \right)^2 +
		\left( \frac{m_{H, S}^2}{m_{H, {\rm exp}}^2} \right)^2 \right\}^{-1/2}
\label{eq:tuningSinglet}
\end{equation}
which is a mild $\Delta \sim 25 \%$ tuning for $1/R \sim 4 \ \TeV$ (while we have fixed $\tan\beta$ for this estimate, additional tuning may be introduced in order to maintain the low $\tan\beta$ needed to lift the physical Higgs mass).

\section{Phenomenology}
\label{sec:pheno}

In this section we discuss the most interesting phenomenological consequences of MNSUSY theories. The leading signature of MNSUSY is the production of sparticles at the LHC and future hadron colliders. In particular, stops should be discovered at LHC13 for the theory to have a level of fine-tuning better than $\sim 10\%$. Because Higgsinos and gauginos are much heavier than 3rd generation sfermions, the usual natural SUSY signatures of $\tilde t$ and $\tilde b$ decaying to on-shell neutralinos and charginos are absent and replaced by 3-body decays. We discuss several cases where extra light states are present in the spectrum and further modify the 3rd generation sfermion decays.  Because there is a large natural hierarchy between the third generation squarks and the rest of the colored superpartners at $1/(2R)$, the latter will only be probed at LHC13 for the lowest scales of $1/R$ compatible with LHC8 results. An approximate KK parity suppresses the single production of the SM KK excitations, so that these states are also likely to be inaccessible at LHC13, though they may have low-energy flavor signals \cite{MNSUSYflavor}. We also briefly discuss further collider signatures of some of the extensions motivated by lifting the Higgs mass.

\subsection{Light States from a Chiral $U(1)_R$}
\label{sec:ChiralSinglets} The models we have studied introduce new sectors related to the Higgs mass and radion stabilization, which contain additional states beyond those minimally related by the 4D and 5D SUSY to the Standard Model particles. New light singlet fermions are particularly relevant for phenomenology, and can generically arise from these sectors if the $U(1)_R$ symmetry is chiral. We will discuss how this occurs in the $U(1)'$ model discussed in Section~\ref{sec:singlet}, the singlet model discussed in Section~\ref{sec:DTerms}, and more generally in simple sequestered models for the extra on-brane SUSY breaking $F_X$. As discussed in Appendix~\ref{app:MaximalRadionMed}, the $U(1)_R$ can be exact or broken by a small amount parameterized as a deviation $\delta\alpha$ from a maximal twist, $\alpha=1/2 + \delta\alpha$. When the $R$-symmetry is broken, the light states can obtain Majorana masses proportional to $\delta\alpha$. These sectors can also be extended beyond their minimal content to be non-chiral at low energies, leading to small Dirac masses for these states. 

\subsubsection*{$U(1)'$ extension}
The  model presented in Section~\ref{sec:DTerms} is chiral under the preserved $U(1)_R$ symmetry, with one linear combination of the $R=-1$ fermions remaining massless, \beq \chi_1 = \cos\theta \bar{\lambda}' + \frac{1}{\sqrt{2}}\sin\theta(\tilde{\phi}_1 - \tilde{\phi}_2). \label{eq:U1LightFermion1}\eeq If the symmetry breaking dynamics were instead localized on the $y=\pi R$ brane giving a vev to $\phi^c_{1,2}$, then the massless state would be \begin{equation} \chi_2 = \cos\theta \lambda' + \frac{1}{\sqrt{2}}\sin\theta(\tilde{\phi}^c_1 - \tilde{\phi}^c_2). \label{eq:U1LightFermion2} \end{equation}
For $m_{Z^\prime} \sim 1/R$, the mixings are large $\theta \sim \mathcal{O}(1)$.
If the $R$-symmetry is broken by a deviation from maximal twist, the state obtains a Majorana mass $\sim \delta \alpha / R$.
The theory can also be made non-chiral by introducing another $R=+1$ fermion coupled through HDOs,
for instance a brane-localized singlet hypermultiplet $V'$ coupling through a \kahler operator
\beq
{\mathcal K} \supset \delta(y) \frac{\tilde g^2}{M_*^3}(V^\dagger V' \Phi_1^\dagger \Phi_1) + {\rm h.c.}
\eeq
which leads to a Dirac mass $M_{\chi}\sim 1/(M_* \pi R)^2 (1/\pi R) \sim~\mathcal{O}(1) \gev$. 

\subsubsection*{Singlet extension}
The model presented in Section~\ref{sec:singlet} also has a chiral spectrum of fermions under the preserved $U(1)_R$, and a state dominantly composed of the $R=+1$ singlet $\tilde{S}$ remains massless, \beq \chi \approx \tilde S + \epsilon (\tilde{h}_d^c \sin\beta + \tilde{h}_u^c \cos\beta ) \label{eq:SingletLightFermion}\eeq with $\epsilon \sim \lambda_S v R \sim 0.1$. Again, this state can be given a Dirac mass $\mu_S$ by introducing an $R=-1$ partner $S'$, coupled for example as $\Delta W = \mu_S S S'$. If $S'$ gets a soft mass ${\tilde m}^2_{S'}$ through HDOs couplings to $F_X$, then the singlet mechanism for the Higgs mass is not spoiled so long as $\mu_S^2 \lesssim {\tilde m}^2_{S'}/10$, which gives $\mu \lesssim 100 \gev$. If the $R$-symmetry is broken by $\delta \alpha$, then the state gets a Majorana mass $\sim \epsilon^2 \delta \alpha / R$.

\subsubsection*{Brane pseudo-goldstino}
For a simple model of a sequestered brane-localized SUSY breaking sector, we take $X$ to be a brane localized singlet with a localized superpotential $\Delta W = \delta(y) \kappa X$ inducing $F_X = \kappa$. There is a light singlet in this model protected by two mechanisms. First, the spectrum is chiral, with $\psi_X$ the only unpaired R-charged state. The gravitino eats a single linear combination of $\psi_X$ and the fermion partner of the radion $\psi_T$ (which plays the role of the goldstino of SSSB \cite{Kaplan:2001cg,Marti:2001iw}), and the uneaten pseudo-goldstino $\chi$ is left massless. For $F_X \sim 1/(\pi R)$, the uneaten light pseudo-goldstino is primarily composed of $\psi_X$ and inherits its goldstino-like couplings to the brane states through the HDOs communicating the extra SUSY breaking, Eq.~(\ref{eq:stopHDO}). The pseudo-goldstino also mixes with the 5D Higgs multiplet through operators of the form of Eq.~(\ref{eq:m2HuHDO}), $\chi \approx \psi_X + \epsilon \tilde{H}_u^c$ with $\epsilon \sim v R  F_X / ( (M_* \pi R) M_*^2) \lesssim 10^{-3}$.  Even in the presence of $R$-symmetry breaking $\delta\alpha$, a combination of $\psi_X$ and $\psi_T$ remains as a massless pseudo-goldstino protected by the special sequestered form of pure brane SUSY breaking in Scherk-Schwarz models~\cite{Cheung:2011jq}. The light state can be given a mass if both (i) the $R$-symmetry is broken or made non-chiral and (ii) SSSB couples directly to the brane SUSY breaking dynamics, spoiling the special brane sequestering. The latter condition will be satisfied in general by loop-level interactions and mixings with the Higgs which couple the brane to the bulk sectors, but the mass of the light state will be suppressed $M\ll1/R$ even in the presence of $\mathcal{O}(1)$ $R$-symmetry breaking.

\subsubsection*{Limits on light states}

A massless gauge singlet fermion is allowed by all astrophysical and collider constraints. Direct production in supernovae and at colliders put limits on the couplings of this state to the Z-boson and on the 4-fermion interactions generated after integrating out the superpartners. These bounds have been studied for example for bino-like states~\cite{NeutralinoMassBounds, NeutralinoSupernovaBounds}, and can easily be satisfied for the candidates discussed above given the high scale $\gtrsim~2~\TeV$ of the gaugino and 1st and 2nd generation superpartners. Assuming a standard thermal history, these 4-fermion operators also determine the decoupling temperature $T_D$ and the corresponding effect on the effective number of neutrino species, which is constrained to be $\Delta N_{eff} \lesssim 0.6$  \cite{NeffPlanck}, requiring $T_D \gtrsim 100~{\rm MeV}$. States with a mixing $\epsilon \lesssim 0.1$ with weak-charged states and four-fermion operators generated from superpartners $\gtrsim 2~{\TeV}$ satisfy this bound (note that four-fermion interactions involving the 3rd generation are unimportant since $T_D \ll m_{\tau,b,t}$), and we find that all of the examples can be made compatible with these limits. 

If the $R$-symmetry is broken to an $R$-parity, or if a new state is introduced to generate a suppressed Dirac mass, then these light states can be cosmologically stable and contribute to the cold dark matter of the universe. The gaugino-like or singlet-like state may obtain a WIMP-like thermal relic density. The pseudo-goldstino, like a light gravitino, may be a viable dark matter candidate for suitable relationships between the reheating temperature and the mass. A detailed exploration of these possibilities is beyond the scope of this work. 

\subsection{Direct Production of 3rd Generation Sfermions}

As discussed in Section~\ref{sec:Fx}, although the overall mass scale of the 3rd generation sfermion masses is set by the 1-loop Scherk-Schwarz scale $\sim 1/(10 R)$, the extra non-predictive sources of SUSY breaking are of a similar size. The free parameters for the phenomenological studies of this model are therefore the relative values of the 3rd generation sfermion masses and the overall scale set by $1/R$. In the $R$-symmetric limit there is no mixing between the right-handed and left-handed sfermions, and the spectrum is therefore completely determined by the soft masses. The strongest limits will typically come from the large pair production cross sections of the 3rd generation squarks, but the other sfermions may be produced in the decays of these states.


\begin{figure}[h]
\begin{centering}
\includegraphics[scale=0.35]{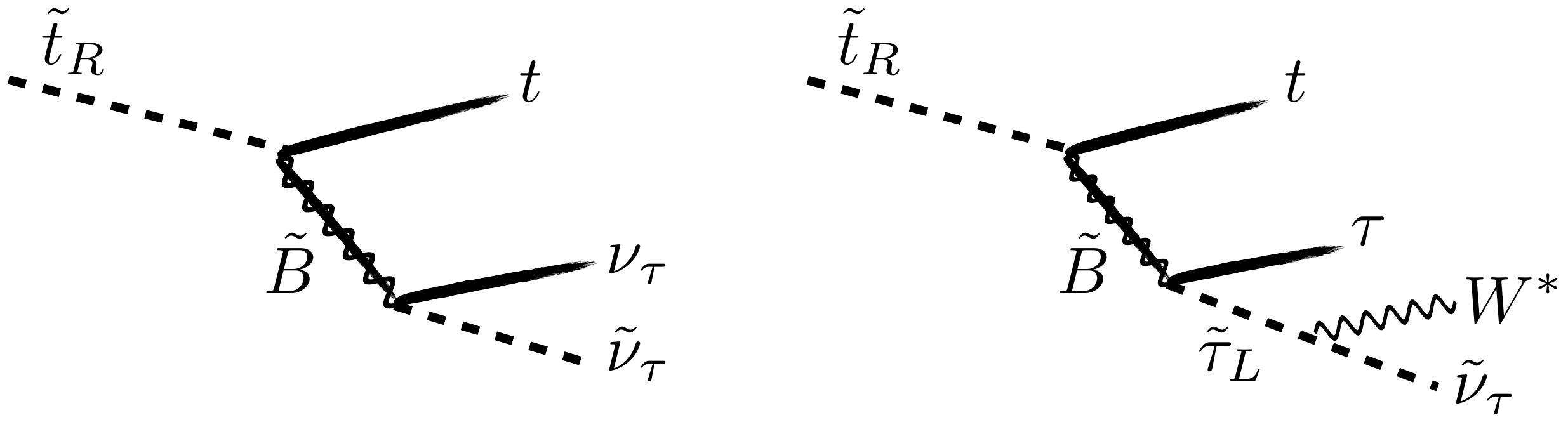}
\caption{\label{fig:ThreeBodyDecay} 
	Decays between the 3rd generation sfermions go through off-shell gauginos and Higgsinos. For example, the decays of $\tilde{t}_R$ to $\tilde{l}=(\tilde{\nu}_\tau,\tilde{\tau}_L)$ are shown.}
\end{centering}
\end{figure}

The decays between third generation states proceed through off-shell gauginos and higgsinos, as depicted in Figure~\ref{fig:ThreeBodyDecay}.  This is an unusual feature of MNSUSY models, as natural spectra in MSSM-like models typically require at least a light Higgsino, and on-shell two-body decays through this state tend to dominate. Because of the Dirac nature of the gaugino and higgsino masses, the three-body decay rate scales as
\beq
\Gamma_{3bdy} \sim \frac{g^4}{192\pi^3}\frac{m^5_{\tilde f}}{1/(2R)^4}. \label{eq:DecayRate3bdy}
\eeq
This rate is prompt for the scales we are interested in, and if no other states are relevant on collider time scales then all decay chains will proceed to the lightest 3rd generation sfermion. If the $\tilde \nu_\tau$ sneutrino is the lightest brane state, then the signatures will involve missing energy. The topology of these decays can nonetheless differ substantially from usual natural SUSY spectra because of the absence of a light Higgsino, with unusual $\tau$-rich final states and diluted missing energy \cite{Alves:2013wra}. If a charged particle is the lightest brane state instead of $\tilde \nu_\tau$, then limits on direct and cascade production of stable charged particles will apply \cite{Aad:2015qfa,Chatrchyan:2013oca} (such a state can be stable on collider time-scales and still decay to satisfy cosmological bounds).


\begin{figure}[h]
\begin{centering}
\includegraphics[scale=0.35]{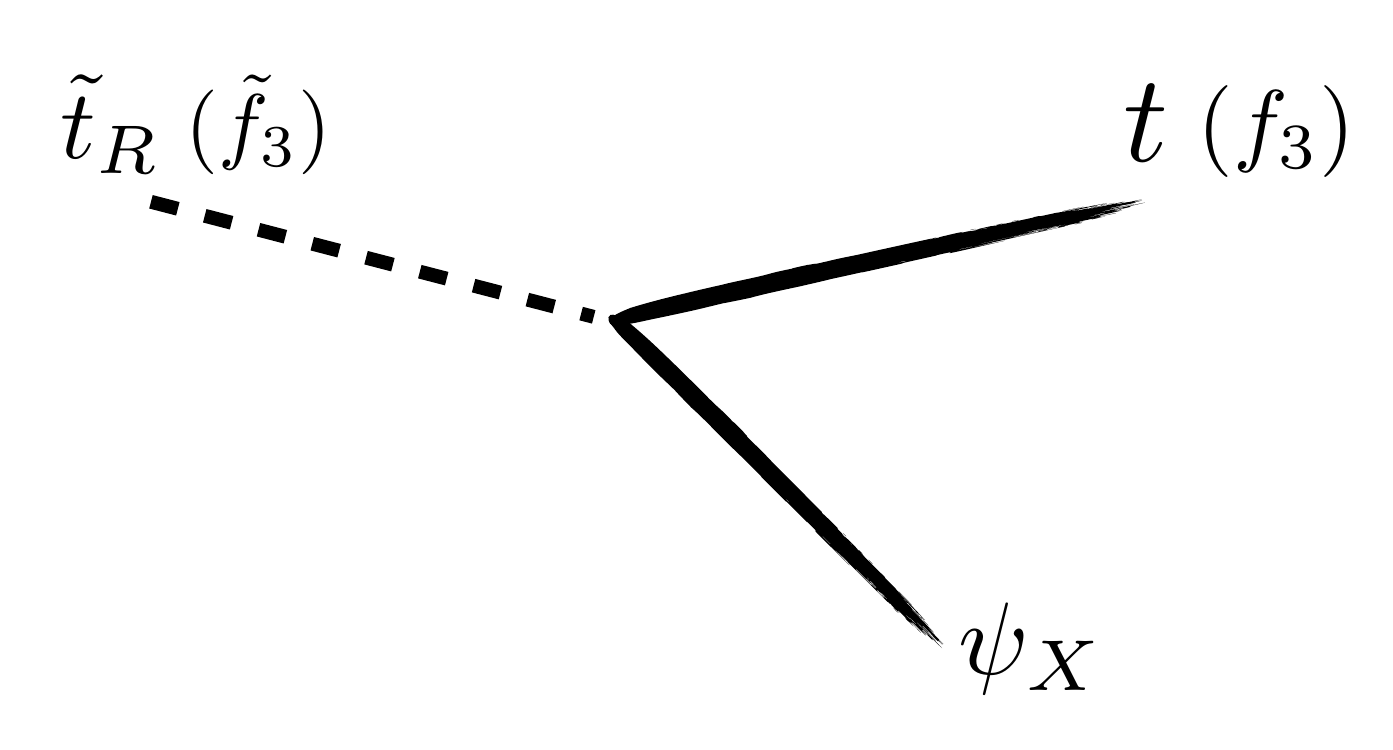}
\caption{\label{fig:GoldstinoDecay} 
	Two-body decays of 3rd generation sfermions ($\tilde{f}_3$) directly to a light pseudo-goldstino LSP
	and a 3rd generation SM fermion ($f_3$)
	can be competitive with or dominate over three-body decays between the 3rd generation states.}
\end{centering}
\end{figure}

\begin{figure}[h]
\begin{centering}
\includegraphics[scale=0.35]{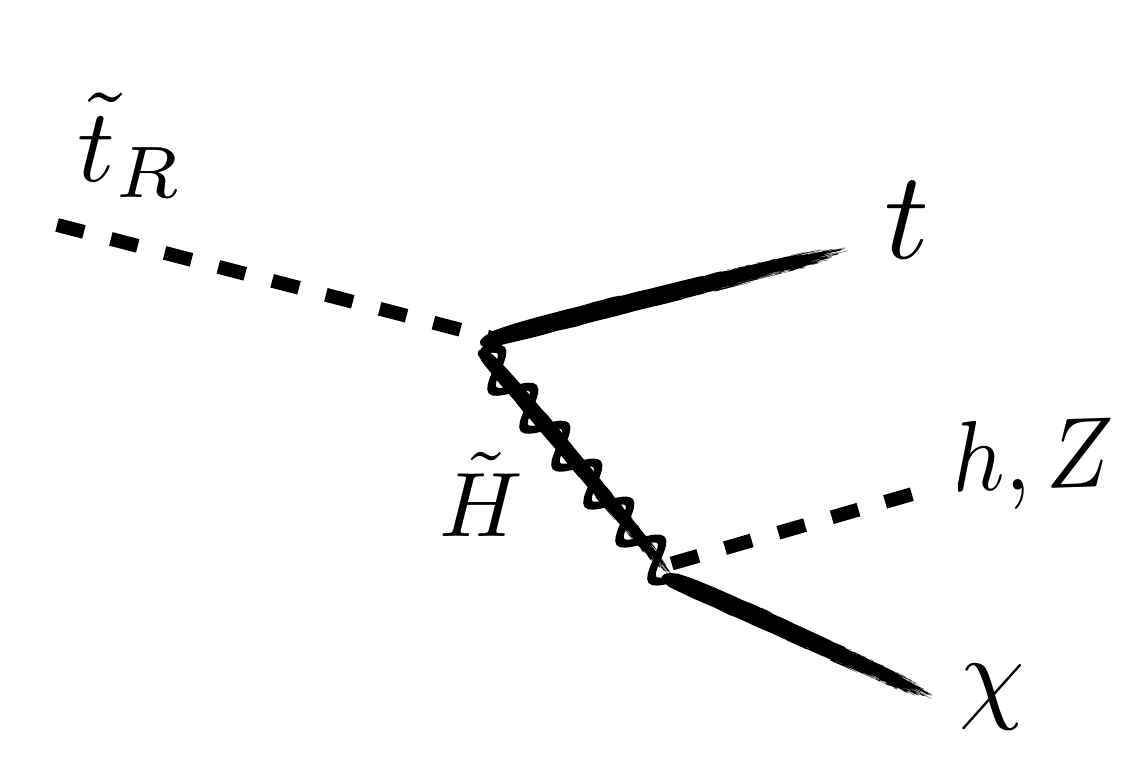}
\caption{\label{fig:ThreeBodyLightDecay} 
	A light LSP protected by a chiral $R$-symmetry may not couple directly to the 3rd generation sfermions. In this case, decays to the LSP will go through off-shell bulk states as shown here. Typically the three-body decays directly to other 3rd generation sfermions depicted in Figure~\ref{fig:ThreeBodyDecay} will dominate, and most decay chains will go first through the lightest 3rd generation sfermion.}
\end{centering}
\end{figure}

If one of the light states described in Section~\ref{sec:ChiralSinglets} is the LSP, then there are new possibilities for the decays of the 3rd generation sparticles. Because the 3-body decay rate Eq.~(\ref{eq:DecayRate3bdy}) is very suppressed, direct decays to these new states can be competitive even when 3-body decays to other 3rd generation sfermions are kinematically accessible. For example, decays to the pseudo-goldstino can completely dominate over 3-body decays, leading to very simple decay topologies directly from 3rd generation states to a massless fermion as depicted in Figure~\ref{fig:GoldstinoDecay}, similar to the MSSM with only a light pure bino LSP accessible. In the opposite limit, when the 3rd generation sfermions do not directly couple to the light states, the normal 3-body decays can dominate and all decay chains pass first through the lightest sfermion. This can be the case for the the singlet model where the light state mixes only ${\tilde h}^u_c$ (see Eq.~(\ref{eq:SingletLightFermion})) or the version of the $U(1)'$ model where the light state mixes only with $\bar{\lambda}'$ (see Eq.~(\ref{eq:U1LightFermion1})), as depicted in Figure~\ref{fig:ThreeBodyLightDecay}. More complicated patterns are also possible, for example if the light fermion in the $U(1)'$ model mixes with $\lambda'$ (Eq.~(\ref{eq:U1LightFermion2})) then the right-handed sfermions can decay directly while the left-handed states must first go through 3-body decays to a right-handed state. All of these possibilities differ substantially from MSSM-like decays to gaugino or gravitino-like LSPs, and are interesting candidates for further study.

Another interesting non-standard possibility is that the 5D MNSUSY model is embedded in a model with large gravitational dimensions. In such a model, the lightest 3rd generation sfermion can decay to states propagating in the large bulk, leading dynamically to signatures similar to compressed spectra. This possibility was studied in detail in Ref.~\cite{Dimopoulos:2014psa}, and can substantially reduce limits on the lightest 3rd generation sfermion. 

In this work we have focused on the phenomenology of the 3rd generation sfermions in the minimal MNSUSY model and its simple extensions related to the Higgs mass.  However, we also note that a number of MSSM extensions focused on modified SUSY signatures that reduce tensions with LHC limits on the lightest colored sparticles could also be incorporated into these models (for example baryonic $R$-parity violation, compressed spectra, stealth SUSY, etc.-- see Ref.~\cite{Evans:2015caa} and references therein). These mechanisms typically only significantly reduce limits on spectra with a very heavy gluino \cite{Evans:2013jna}, and therefore the natural hierarchy between the gluino and the 3rd generation sfermions in MNSUSY models offers an appealing framework for these models. 

\subsection{Bulk states}

While the leading signature of supersymmetry in MNSUSY is the production of the 3rd generation states, the presence of \emph{near-degenerate} first and second generation sfermions and gauginos along with their $\mathcal{N}=2$ counterparts at $1/(2R)$ would be a strong indication of the intrinsically extra-dimensional nature of the SUSY breaking that may be within LHC13 discovery reach for $1/R \lesssim 6~\TeV$ or within reach of a $100~\TeV$ proton collider for $1/R\lesssim 30~\TeV$ \cite{Cohen:2013xda}.

Another probe of the extra-dimensional nature of the MNSUSY model is the production of the SM KK excitations. Because the first two generations propagate in the bulk, there is an approximate KK-parity and a suppression of the single-production of the 1st KK resonances. If brane-localized kinetic terms (see Eq.~(\ref{eq:BraneLocalizedKinetic})) of NDA-size are present and fully violate KK-parity, single-production cross-sections of KK-gauge bosons at $1/R$ are suppressed by a factor $\sim 0.01-0.05$. Couplings to the brane-localized third generation on the other hand have no suppression, and the decays of these resonances will be dominantly to the third generation. Current LHC limits on a KK $Z$ with such a suppressed cross section decaying dominantly to 3rd generation states are $m_{Z'} \gtrsim 1.5 \tev$ \cite{Aad:2015osa}, and there is comparable mass reach for a KK gluon decaying to 3rd generation quarks \cite{ATLAS-CONF-2013-052, Datta:2013lja,Khachatryan:2015sma}.  Even the $13~\TeV$ LHC run is unlikely to probe into the most interesting range $1/R \gtrsim 4~\TeV$, but if sparticles are discovered at LHC13, the first KK resonances are likely to be accessible at a $\sim 100~\TeV$ proton collider. We also note that there is no approximate KK-parity in variants of the model where all three generations are localized on the brane, in which case single production of KK resonances can be among the leading signatures.

Although direct production of KK states is not a strong signal in MNSUSY, the exchange of KK gauge bosons generates contributions to flavor changing neutral currents (FCNC). The non-minimal flavor structure due to the localization of the 3rd generation leads to signals in processes involving 3rd generation fermions (e.g. $B$-meson mixing). As discussed in detail in Ref.~\cite{MNSUSYflavor}, flavor violation is expected at interesting levels, although variants of the model can suppress these signatures, for example by allowing some of the 3rd generation matter to propagate in the bulk.

A last probe of the 5D nature of the model is the presence of a light radion parameterizing the size of the extra dimension. The mass of such a state depends on the nature of the embedding of the 5D theory in a full gravitational theory explaining $M_{\rm Planck} \gg M_*$, and can be as light as $\sim 1/(R^2 M_{\rm Planck})$. This state is potentially accessible in equivalence principle and fifth force tests \cite{Antoniadis:1997zg}.

\subsection{Vector-like Leptons}
\label{sec:phenoVL}

If the $\z$ symmetry acting on the VL-leptons of Section \ref{sec:VL} were exact, this version of the model would already be ruled out.
Current constraints from LHC data seem to suggest that the possibility of a stable charged particle with a mass below $\sim 340$ GeV is experimentally
excluded \cite {CMSStableChargedParticles} and, for some reasonable values of our parameters, the mass of our lightest VL-lepton is $m_{\tau_1^\prime} \approx 260 \ \GeV$.
For this reason, it is convenient to assume that the $\z$ symmetry is in fact broken by HDOs that
allow mixing between the VL and SM sectors. For simplicity, we will assume mixing only happens with the 3rd generation of SM leptons, so that
the lightest VL state, $\tau_1^\prime$, will decay to the $\tau$ sector, whereas we expect that the other two VL states, $\nu^\prime$ and $\tau_2^\prime$,
would dominantly decay to $\tau_1^\prime$ via interactions that preserve the $\z$ symmetry.
In particular, \kahler operators of the form
\begin{equation}
	\mathcal{K} \supset \delta(y) \frac{{\tilde g}^2 {\tilde k}^\prime}{\sqrt{3 \pi / 2} M_*^{5/2}} X^\dagger H_u^\dagger (L_3 \tilde L + \tilde L {\bar E}_3)
\label{eq:VLmixingHDOs}
\end{equation}
would result in the decay $\tau_1^\prime \rightarrow h \tau$, characterized by a 4D effective Yukawa coupling
$k^\prime = 4 \cdot 10^{-3} {\tilde k}^\prime f_X (25 / (M_* \pi R))^{3/2}$.
For the decay to occur promptly on collider timescales, the corresponding decay rate $\Gamma$ must satisfy $\Gamma^{-1} \lesssim 100 \ {\rm \mu m}$, which is
satisfied for a $\z$-breaking Yukawa $k^\prime \gtrsim 10^{-6}$.
Smaller values of $k^\prime$ would result in displaced vertices at colliders, and the decay $\tau_1^\prime \rightarrow h \tau$ remains cosmologically
safe (i.e. $\Gamma^{-1} \ll 1 \ {\rm s}$) so long as $k^\prime \gtrsim 10^{-12}$.

Although, to the best of our knowledge, a dedicated search for VL-leptons decaying to SM states using LHC data is not available,
some results in the literature seem to suggest that for a lightest VL-lepton decaying to the $\tau$ sector, a lower limit on its mass may be
as high as $270$ GeV if that state is mostly $SU(2)_L$ doublet and below $100$ GeV if mostly singlet \cite{VLleptonsFalkowski}.
In our case, $\tau_1^\prime$ is, to a very good approximation, an even mix of doublet and singlet and therefore we expect that its mass
being $260$ GeV is allowed by current data. Albeit a dedicated study would be needed, it seems that with such low mass, the lightest
VL-lepton of our model may be within reach of being discovered at LHC, with the possibility of displaced vertices in a significant range of the allowed
values of the $\mathbb{Z}_2$ breaking Yukawa.

\section{Conclusions}
\label{sec:conclusions}

Since 1981 the search for a successful SUSY generalisation of the SM has been
dominated by the paradigm of the MSSM and its many variants.    Essentially all models
of weak scale SUSY assume, as a very first step, that the theory can be adequately
described as a softly broken SUSY theory in 4D, more particularly a $\mathcal{N}=1$ 4D
SUSY theory with some yet-to-be determined dynamics that sets the structure of the soft terms. 
Even though this paradigm has attractive features, particularly the successes of SUSY gauge
coupling unification and radiative EWSB, since LEPII it has been become increasingly apparent
that either some structural feature is lacking in our implementation of SUSY as it applies to weak scale physics or naturalness may not be a reliable guide to new physics at the weak scale.

With the discovery of the Higgs, but the lack of any experimental sign of superpartners at LHC run 1, this dilemma has sharpened --- the Higgs
is seemingly well-described as an elementary scalar, but the expected (colored) superpartners which, according to the MSSM
must be present at accessible energy scales for a natural weak scale, are apparently absent.
Generally,  the question we face is if we should give up the idea of naturalness as it is currently understood, or should we search for
qualitatively new implementations of natural theories.   

In this work we take the concept of naturalness as a good guide to constructing the correct theory of the weak scale,  demanding that the
weak scale has a dynamical explanation. On the other hand we are willing both to move away from the structures of the MSSM and its variants,
and also temporarily give up the ambitious project of explaining the entire $M_{\rm Planck}$ to $M_W$ hierarchy.  Instead we focus on a natural, dynamical, explanation of the little hierarchy problem in a framework that, we believe, is ultimately extendable to addressing the full hierarchy problem.  

Maximally Natural Supersymmetry (MNSUSY) achieves these aims, providing a calculable existence proof that natural SUSY
theories of the weak scale are possible, although they may differ substantially from the softly broken MSSM. Moreover we have shown that, compatible
with all current experimental constraints, these theories have a remarkably low level of tuning relative to conventional
theories, and they can lead to unusual and striking signatures at colliders.

The crucial ingredient behind the success of MNSUSY is the Scherk-Schwarz mechanism (with maximal twist) of SUSY breaking in 5D.
The non-local nature of this breaking ensures that SUSY breaking parameters are only sensitive to scales up to the
compactification scale $1/R$ ($\gtrsim 4 \tev$ satisfies all constraints), and are insensitive to the UV cutoff, even at the logarithmic level.
The 5D geography of fields also plays a major role:
whereas gauge and Higgs sectors propagate in the 5D bulk (often, but not absolutely necessarily, together with the 1st and
2nd matter generations), the 3rd generation remains localized on one of the branes.
These two features act together leading to a 4D effective theory where the usual problems of SUSY theories are solved (for example
the $\mu$ and $B_\mu$ problems, and the problem of the radiative sensitivity of EWSB to the gluino mass) with a low level of tuning.

The minimal implementation of the model predicts, however, a Higgs mass $m_h < 125 \gev$
if the theory is restricted to the low fine-tuning region.
Simple extensions of the minimal theory solve this problem and do not significantly affect the physics of EWSB.
We have explicitly discussed three qualitatively different extensions:
the presence of a family of brane-localized vector-like leptons that couple to the Higgs with an $\mathcal{O}(1)$ Yukawa coupling;
the addition of extra $U(1)$ gauge structure under which the Higgs is charged;
and an NMSSM-like extension where both Higgs fields get a non-zero vev and the presence of a brane-localized singlet chiral superfield provides an additional tree-level contribution to the Higgs mass.
We have shown that all of these extensions preserve the qualitatively attractive features of the model and give a successful theory of the weak scale with a level of fine-tuning that is $\sim10 \%$ or better.

\section*{Acknowledgments}

We thank M. Baryakhtar, J. Huang, J. Scoville, T. Cohen, D. Pinner and, especially, Savas Dimopoulos for discussions.
The authors thank the CERN Theory Group and (IGG and JMR) the Stanford Institute for Theoretical Physics
for hospitality during portions of this work.
IGG is financially supported by the STFC/EPSRC and a Scatcherd European Scholarship from the University of Oxford. KH acknowledges support from NSF grant PHY-1316706 and an NSF Graduate Research Fellowship under
Grant number DGE-0645962. Fermilab is operated by Fermi Research Alliance, LLC  under Contract No. DE-AC02-07CH11359 with the United States Department of Energy.

\appendix
\section{A Pirate's Favorite Symmetry}
\label{app:MaximalRadionMed}

Results related to the SUGRA embedding of the Scherk-Schwarz twist and the radion stabilization sector  have usually been studied in the radion mediation picture at small values of the twist parameter \cite{Rattazzi:2003rj,Luty:2002hj,Marti:2001iw,Kaplan:2001cg}.
In this appendix we clarify how these results apply to the case of maximal twist, in particular focusing on the status of the global $U(1)_R$ symmetry present at maximal twist.

\subsection{$U(1)_R$ symmetry in the radion mediation picture}
\label{app:bcpicture}

SSSB at maximal twist has at the global level an exact $U(1)$ $R$-symmetry.
In this section, we clarify how this symmetry is realized in the radion-mediation picture. It is well known that a Scherk-Schwarz twist is dual to radion mediation,
with $F_T = 2 \alpha$ when the radion is normalized to $\langle T \rangle = R$
\cite{Ferrara:1988jx,Porrati:1989jk,Luty:2002hj,Marti:2001iw,Kaplan:2001cg},
and the twist parameter $\alpha$ corresponding to a maximal twist when $\alpha = 1/2$.
For simplicity, we focus on the case of a bulk hypermultiplet with 4D Lagrangian \cite{Marti:2001iw}
\begin{equation}
\mathcal{L}_4 =  \int d{\tilde y} \left\{ \int d^4\theta (\Phi^\dagger \Phi + \Phi^{c\dagger} \Phi^c) \frac{T+T^\dagger}{2}
			+  \int d^2\theta ( \Phi \partial_{\tilde y} \Phi^c + {\rm h.c.} ) \right\}
\end{equation}
with $\tilde y=[0,\pi]$. 

In the twist picture, the parameter $\alpha$ can be viewed as controlling the bc's of the fields at $\tilde y=\pi$.
The ${\mathcal N}=1$ SUSY preserved at $\tilde y=0$ acts on the chiral multiplets $\Phi=(\phi, \psi), \Phi^c = (\phi^c, \psi^c)$;
the ${\mathcal N}=1'$ preserved at $\tilde y=\pi$ acts on the 4D chiral multiplets obtained by an $R$-symmetry rotation,
$\Phi' = (\phi', \psi)$, ${\Phi^c}' = ({\phi^c}', \psi^c)$, with the new scalars defined as 
\begin{equation}
(\phi'\;\; {\phi^{c \prime}}^\dagger) = e^{i \sigma_2 \pi \alpha}(\phi\;\; {\phi^{c\dagger}}).
\end{equation}
A choice of bc's locally conserving the separate ${\mathcal N}=1,1'$ supersymmetries and leaving a fermion 0-mode is simply 
\begin{equation}
\Phi^c|_{\tilde y=0}=0, \;\; {\Phi^c}'|_{\tilde y=\pi}=0
\label{eq:twistedBC}
\end{equation}
The choice of bc's at $\tilde y=0$ is compatible with the $U(1)_R$ subgroup of $SU(2)_R$ with elements
$U(\beta)=e^{i \sigma_3 \beta}$.
For generic values of twist $\alpha$, this symmetry is not compatible with the bc's at $\tilde y=\pi$.
However, at $\alpha=1/2$, the $R$-symmetry is preserved at both boundaries, resulting in the theory at maximal twist having broken SUSY
and an exact global $U(1)_R$ symmetry. 

To go to the Wilson line frame of Scherk-Schwarz, a non-periodic $SU(2)_R$ gauge transformation can be performed \cite{vonGersdorff:2001ak}.
In the radion effective theory, this is simply a $\tilde y$-dependent field redefinition 
\begin{equation}
(\phi\;\; {\phi^{c\dagger}}) \rightarrow e^{i \sigma_2 \alpha \tilde y}(\phi\;\; {\phi^{c\dagger}}),
\label{eq:Rtwist}
\end{equation}
with the extra part of the 5D derivatives on the new fields absorbed in
\begin{equation}
T' = T + 2 \alpha \theta^2
\label{eq:FTredefinition}.
\end{equation}

The bc's of the redefined fields are consistent with a single orbifold preserving the same ${\mathcal N}=1$ SUSY,
and the remaining SUSY breaking is parameterized completely as radion mediation through the  non-vanishing $F_T'$.
The gravitational part of the 4D radion effective action \cite{Luty:2002hj} in terms of the shifted $T'$ of course contains
a suitable superpotential term to stabilize $F_T' = 2 \alpha$, 
\begin{equation}
- \int d^4\theta 3 M_5^3 \phi^\dagger \phi \frac{T+T^\dagger}{2} \rightarrow - \int d^4\theta 3 M_5^3 \phi^\dagger \phi (\frac{T'+{T'}^\dagger}{2}) +
\left( \int d^2 \theta \phi^3 \alpha M^3_5 + {\rm h.c.} \right) , \label{eq:Tnoscale}
\end{equation}
where the fixing of the conformal compensator $\phi = 1 + F_\phi \theta^2$ has been used to rewrite the constant superpotential in the usual form \cite{Luty:2002hj}, and where $M_5$ is the gravitational scale, which may be the same size as or parametrically larger than the fundamental 5D scale $M_*$ depending on the UV completion.
While the existence of an $R$-symmetry at $F_T=0$ is clear in the radion mediation picture, the existence of a restored $R$-symmetry at the maximal value
$F_T=1 $ is unclear until the spectrum is calculated -- naively the fact that $\langle T \rangle = R \neq 0$ and $\langle F_T \rangle = 2 \alpha \neq 0$ seems to preclude the definition of a conserved $R$-symmetry.
However, we are interested in an $R$-symmetry that is global from the 4D perspective, but may be $\tilde y$-dependent in 5D.
The non-linear transformations of the derivative terms under such a symmetry can then cancel the variation of the action under
the non-linear transformations of $\langle F_T \rangle$. The suitable symmetry can be inferred from the field redefinition Eq.~(\ref{eq:Rtwist}), giving 
\begin{equation}
U(y,\theta) = e^{i\sigma_2 \frac{1}{2} \tilde y}e^{i\sigma_3 \theta}e^{-i \sigma_2 \frac{1}{2} \tilde y}.
\end{equation}
As in the twist picture, this is only a (4D) global symmetry of the theory at the maximal twist, $F_T = 2 \alpha = 1$.

\subsection{$U(1)_R$ symmetry with radius stabilization and 5D curvature}

In flat Scherk-Schwarz theories without radion stabilization, the 4D radion effective action has the no-scale form, and thus no vev is generated at tree level for the conformal compensator. When the radion is stabilized, the no-scale form is broken and for general values of the twist a nonzero $F_\phi$ is generated \cite{Luty:2002hj},  giving an extra source of SUSY breaking and $R$-symmetry breaking. As we review, even for general twist $F_\phi$ is parametrically smaller in the flat case than other sources of SUSY breaking. Furthermore, we show that when the twist is maximal and all brane-localized sectors also preserve the $U(1)_R$ symmetry, their contributions to the radion potential can not generate a nonzero $F_\phi$. SUGRA effects at non-vanishing 5D curvature can also preserve the $R$-symmetry.

We work in the frame where the bc's correspond to a twist $\alpha$, and where $F_T$ parameterizes shifts away from an $\alpha$ twist. Integrating out the matter sector generates Casimir energies, as well as tree-level terms if there are non-trivial bulk-brane dynamics, all of which depend on the twist $\alpha + F_T/2$. These can be parameterized in a potential $V(\alpha + F_T/2, R)$, which we assume leads to stabilization of the radius.
As before, we assume that the scale of $V$ is set by the Casimir energies, $V\sim 1/(\pi R)^4$. With the tree-level gravitational action, this gives
\begin{equation}
\mathcal{L}_4 \supset - \int d^4\theta \left[ 3 M_5^3 \phi^\dagger \phi \frac{T+T^\dagger}{2} \right] - V|_{F_T=0} - \frac{\partial V}{\partial F_T}|_{F_T=0} F_T + ...~.
\end{equation}
For generic $\mathcal{O}(1)$ values of $\alpha$, $\frac{\partial V}{\partial F_T}\sim  1/(\pi R)^4$ and solving the $F$-term equations of motion yields $$F_\phi \sim \frac{1}{M_5^3}\frac{\partial V}{\partial F_T} \sim  \frac{1}{\pi R}\frac{1}{(\pi M_5 R)^3}.$$ Therefore the SUSY breaking communicated by $F_\phi$ is heavily suppressed compared to the SUSY breaking communicated by $F_T$. Higher order terms in $F_T$ do not affect this conclusion. 

In the case of maximal twist, $\alpha = 1/2$, the value of $F_\phi$ can be even further suppressed. $F_T$ has $Q_R = -2$ under the $U(1)_R$ symmetry preserved at maximal twist. If the brane-dynamics also preserves the $R$-symmetry (as for example in the models described in this work), then integrating out the matter fields will lead to an $R$-symmetric potential, and the linear term must vanish, $\frac{\partial V}{\partial F_T}=0$. This leads to $F_\phi = 0$, and the conclusion is again insensitive to the higher order terms in $F_T$ allowed by the $R$-symmetry.

The flat 5D SUGRA has a global $SU(2)_R$ symmetry, and introducing a 5D CC (giving 5D curvature $k=\sqrt{-\Lambda_5/M_5^3}$) breaks the global symmetry down to a $U(1)_R$ \cite{Fujita:2001bd}. General choices of bc's on the branes break the remaining global $U(1)_R$ symmetry. Theories with a 4D SUSY preserved by the bc's are the ``detuned" models of~\cite{Bagger:2002rw}, and the $R$-symmetry violation is evident in the 4D theory from the non-vanishing 4D supersymmetric CC. The maximal bc's however break SUSY while preserving the $R$-symmetry at both
branes\footnote{$\alpha_{0} = \infty, \alpha_{\pi}=0$ in the notation of Refs.~\cite{Bagger:2002rw}. For non-vanishing bulk curvature, this requires a non-vanishing source of spontaneous SUSY breaking $F_X^2\sim k M_5^3$ on the $y=0$ brane to obtain a vanishing 4D CC \cite{Bagger:2004rr}.
} 
\cite{Bagger:2002rw,Gherghetta:2000kr}.  Thus although non-vanishing $k$ can perturb the spectrum, we note the choice of maximal bc's still preserves the global $U(1)_R$ symmetry in this case and does not generate a nonzero $F_\phi$. In the models we consider there is no need for $k$ to be related to the scales $M_*$ or $1/R$, and we have focused on the flat bulk case where $k \ll R^{-1}$. 

While the maximal bc's correspond to an enhanced $U(1)_R$ global symmetry and are therefore technically natural, one might expect non-perturbative string effects to break this global symmetry. The leading effect would be a small shift $\delta \alpha$ away from maximal bc's. This will introduce $R$-parity effects directly into the bulk spectrum, and as a subdominant effect induce $F_\phi \sim  \delta \alpha \times 1 / ( \pi R (M_5 \pi R)^3 )$. We are therefore justified in ignoring the effects of $F_\phi$, and we generally assume $\delta\alpha \ll 1$. Non-perturbative breaking of the $R$-symmetry could also induce suppressed HDOs violating the $R$-symmetry on the brane.

\subsection{$U(1)_R$ symmetry and brane-localized masses}

Scherk-Schwarz bc's of the form Eq.~(\ref{eq:twistedBC}) can also be obtained from a ${\mathcal N}=1$ SUSY preserving orbifold by localizing SUSY-breaking $R$-symmetry violating mass terms for hypermultiplets ($\Phi$) and gauginos ($\lambda$) on the $\tilde y=\pi$ brane,
\beq
\mathcal{L_{\rm BC}}=\delta(\tilde y-\pi) (\tilde m \lambda \lambda  + \tilde m \phi F^\dagger_\Phi + {\rm h.c.})
\eeq
These mass terms generate jumping profiles for the odd fields $\lambda^c, \phi^c$ over the brane, giving twisted bc's on the interval with the correspondence $\tilde{m}=-2\cot(\pi\alpha)$ \cite{Bagger:2001qi}. While these boundary terms violate the ${\mathcal N}=1$ SUSY preserved by the orbifold, a new ${\mathcal N}=1$ SUSY transformation can be defined under which the odd fields and even fields are mixed by a jumping profile -- this ${\mathcal N}=1'$ SUSY is broken at $\tilde y=0$ but preserved at $\tilde y=\pi$, and thus all of the features of Scherk-Schwarz are recovered \cite{Bagger:2001qi,Bagger:2001ep,Delgado:2002xf}.

In this picture, the brane-localized mass term appears to violate the $R$-symmetry strongly, even in the limit that $\tilde m \rightarrow \infty$, corresponding to maximal bc's. How is the $R$-symmetry preserved? As $\tilde m \rightarrow \infty$, the profiles of the low energy states become suppresed near the brane, and the states with unsuppressed $\lambda|_{\tilde y=\pi}$ feeling the $R$-symmetry violation decouple to high energy. On-shell, the equations of motion give $\delta(\tilde y-\pi) \tilde{m} \lambda \lambda = \delta(\tilde y-\pi) (4/\tilde m) \lambda^c \lambda^c$ for the boundary action, and as $\tilde m \rightarrow \infty$ the variation in the action under an $R$-symmetry transformation vanishes. In a deconstructed model this is analagous to decoupling one of the states on the end-sites \cite{Falkowski:2002gx}, leaving a theory with a different chirality and a restored $R$-symmetry.

This picture of Scherk-Schwarz has features in common with theories of gaugino mediation \cite{Kaplan:1999ac,Chacko:1999mi}, where SUSY is completely broken on the $\tilde y=\pi$ brane and is  communicated through gauginos and gravitational states in the bulk. At the level of bc's, this scenario differs from Scherk-Schwarz because the brane-masses need not be equal for all the bulk states, leading to no consistent definition of a surviving supersymmetry on the $\tilde y=\pi$ brane. The softness of the SUSY breaking effects communicated to the $y=0$ brane is similar to Scherk-Schwarz, but bulk scalars can not be protected because of the absence of any preserved SUSY on the $\tilde y=\pi$ brane. SUSY breaking on the $\tilde y=\pi$ brane could be hard or take a normal form for spontaneous $F$- or $D$-term SUSY breaking. In the latter case, typical models realize spectra similar to Scherk-Schwarz at small twists; the large $\tilde m$ limit corresponding to maximal twist involves large $F$-terms and leads to competing anomaly mediated and radion mediated effects.

\footnotesize
\bibliography{MNSUSYsimplifiedRefs}
\end{document}